\documentclass[a4paper,12pt]{amsart}

\usepackage[margin=1in]{geometry}

\usepackage{amsmath, amsthm, amssymb}
\usepackage{stmaryrd}
\usepackage{algorithm,  algpseudocode} 
\usepackage{cite}
\usepackage{graphicx}
\usepackage{datetime2} 

\theoremstyle{definition}
\newtheorem{theorem}{Theorem}[section]
\newtheorem{lemma}[theorem]{Lemma}
\newtheorem{proposition}[theorem]{Proposition}
\newtheorem{fact}[theorem]{Fact} 
\newtheorem{observation}[theorem]{Observation}

\theoremstyle{definition}
\newtheorem{definition}[theorem]{Definition}

\newtheorem{pclaim}[theorem]{Claim}

\theoremstyle{remark}

\newenvironment{rmenum}{
\begin{enumerate}

}
{\end{enumerate}}

\newcommand{\parNei}[2]{N_{#1}(#2)}

\newcommand{\distgt}[4]{\lambda_{(#1, #2)}(#3, #4)}

\newcommand{\distgtf}[5]{\lambda_{(#1, #2)}(#4, #5; #3)}
\newcommand{\distpf}[4]{\lambda_{#1}(#3, #4; #2)}
\newcommand{\distp}[3]{\lambda_{#1}(#2, #3)}
\newcommand{\parcut}[2]{\delta_{#1}(#2)}

\newcommand{\conn}[1]{\mathcal{C}(#1)}

\newcommand{\laylegtr}[4]{L_{(#1, #2)}(#4; #3)}

\newcommand{\levelgtr}[4]{U_{(#1, #2)}(#4; #3)}

\newcommand{\laycompgtr}[4]{\mathcal{L}_{(#1, #2)}(#4; #3)}
\newcommand{\laycompall}[3]{\mathcal{L}_{(#1, #2)}(#3)}
\newcommand{\qcompgtr}[4]{\mathcal{Q}_{(#1, #2)}(#4; #3)}
\newcommand{\qcompall}[3]{\mathcal{Q}_{(#1, #2)}(#3)}

\newcommand{\agtr}[3]{A_{(#1, #2)}(#3)}

\newcommand{\dgtr}[3]{D_{(#1, #2)}(#3)}

\newcommand{\initialgtr}[3]{K_{(#1, #2)}(#3)} 
\newcommand{\qinitialgtr}[3]{K^{\circ}_{(#1, #2)}(#3)}

\newcommand{\interval}[3]{I_{(#1, #2)}(#3)}

\newcommand{\ak}[3]{{A}_{(#1, #2)}(#3)} 
 
\newcommand{\dk}[3]{{D}_{(#1, #2)}(#3)} 
\newcommand{\dkconn}[3]{{\mathcal{D}}_{(#1, #2)}(#3)} 
\newcommand{\qkconn}[3]{{\mathcal{Q}}_{(#1, #2)}(#3)}

\newcommand{\noncapall}[3]{\mathcal{L}^{*}_{(#1, #2)}(#3)}
\newcommand{\qnoncapall}[3]{\mathcal{Q}^{*}_{(#1, #2)}(#3)}

\newcommand{\alpart}[1]{\mathcal{A}_{\mathcal{L}}(#1)}  
\newcommand{\aqpart}[1]{\mathcal{A}_{\mathcal{Q}}(#1)}

\title[Connected Minimum Join]{Constructive Characterization and Recognition Algorithm for Grafts with a Connected Minimum Join}
\author{Nanao Kita}
\address{Nagoya University, 464-8602 Furocho, Chikusa, Nagoya, Japan}
\email{kita at math nagoya-u ac jp}

\begin{document}

\begin{abstract}
Minimum joins in a graft $(G, T)$, also known as  minimum $T$-joins of a graph $G$, are said to be connected if they determine a connected subgraph of $G$. 
Grafts with a connected minimum join 
have gained interest ever since Korach and Penn showed that  
they satisfy Seymour's min-max formula for joins and $T$-cut packings; that is, in such grafts, 
the size of a minimum join is equal to the size of a maximum packing of $T$-cuts.  
In this paper,  we provide a constructive characterization of grafts with a connected minimum join. 
We also obtain a polynomial-time algorithm that decides whether a given graft has a connected minimum join and, if so, outputs one.  
Our algorithm has two bottlenecks; 
one is the time required to compute a minimum join of a graft, 
and the other is the time required to solve the single-source all-sink shortest path problem 
in a graph with conservative $\pm 1$-valued  edge weights. 
Thus, our algorithm runs in $O(n(m + n\log n) )$ time. 
In the nondense case, it improves upon the $O(n^3)$ time bound for this problem, due to Seb\H{o} and Tannier, which was introduced as an application of their results on metrics on graphs. 
\end{abstract}

\maketitle

\section{Introduction}

The minimum $T$-join problem on graphs, also known as the minimum join problem on grafts, is a fundamental problem in combinatorial optimization 
that has been studied since the 1960s~\cite{lp1986, schrijver2003, kv2008, cook1998combinatorial}.  
Given a graph $G$ and a set $T$ of vertices, a set of edges $F$ is called a {\em $T$-join} of $G$, 
or a {\em join} of the pair $(G, T)$, 
if a vertex is incident to an odd number of edges in $F$ if and only if it is in $T$. 
A pair $(G, T)$ of a graph $G$ and a set $T$ of vertices has a join if and only if it is a {\em graft}, 
that is,  every connected component has an even number of vertices in $T$.    
The minimum join problem on a graft is a classical problem in P 
that encompasses several special cases 
such as the undirected shortest path problem, the postman problem, and perfect matching.

Connected minimum joins (CMJs) have attracted attention ever since Korach and Penn~\cite{korach1992tight} discovered that 
grafts with a CMJ satisfy Seymour's min-max formula between $T$-joins and {\em $T$-cut packings}. 
Seymour~\cite{seymour1981odd, lp1986, schrijver2003, cook1998combinatorial} provided the celebrated duality theorem 
 which states that the size of a minimum $T$-join equals the size of a maximum $T$-cut packing in bipartite graphs. 
As duality theorems are guiding principles in combinatorial optimization, 
this opens up the quest for the class of grafts beyond bipartite grafts 
for which Seymour's min-max formula holds. 
Korach and Penn~\cite{korach1992tight} later discovered that, 
if a graft has a {\em connected} minimum join, that is, a minimum join inducing a connected subgraph,  
then it satisfies Seymour's min-max formula. 
Since then, CMJs have gained attention, 
and Frank mentioned in his survey article~\cite{frank1996survey} 
that whether recognizing grafts with a CMJ is in class P remained an open problem.   
In recent years, grafts with a CMJ have been used to provide a tractable class of instances for NP-hard problems, such as the shortest odd path problem~\cite{JUTTNER202434, doi:10.1137/23M158156X}. 
These results cover the shortest odd path problem with $\pm 1$ conservative edge weights, 
although they handled the case of more general conservative edge weights. 

Seb\H{o} and Tannier~\cite{10.1007/3-540-45535-3_30, sebHo2004metric} gave the first polynomial-time algorithm for deciding whether a given graft has a CMJ  
as an application of their results on metric spaces. 
Their algorithm runs in $O(n^3)$ time and  also computes a CMJ in the same time bound if one exists. 
Their approach proceeds as follows. 
They first formulated the isometry extension problem: 
Given metric spaces $S_1$ and $S_2$, a subspace $S_1' \subseteq S_1$,  
and an isometry $f: S_1' \rightarrow S_2$, 
the problem asks whether there exists an isometry $\bar{f}: S_1 \rightarrow S_2$ extending $f$. 
They then identified a class of instances for which this problem can be solved in polynomial time. 
They then proved that the recognition problem for grafts with a CMJ can be reduced to 
this solvable case of the isometry extension problem. 
Their seminal paper has sparked a new line of research on strong metric generators.

In this paper, we first give a constructive characterization of grafts with a CMJ.   
According to our characterization, it turns out that grafts with a CMJ have a very limited structure. 
Specifically, every graft with a CMJ is a recursive combination of a special class of grafts called {\em rakes}, 
which have a CMJ inducing a star, 
with an attached ``tail'', an arbitrary structure that does not affect the minimum join property of the graft.   
Rakes have a very simple structure, and we also give a full constructive characterization of rakes from the ground up. 
Thus, we obtain a full constructive characterization of grafts with a CMJ from the ground up.

We derive our results by graph-theoretic arguments in join theory, 
and the only preliminary result we employ is Seb\H{o}'s distance decomposition of grafts~\cite{sebo1990}. 
This result was not employed in the work of Seb\H{o} and Tannier~\cite{10.1007/3-540-45535-3_30, sebHo2004metric}. 
Seb\H{o}'s distance decomposition is a fundamental tool in the study of joins. 
Given a minimum join, a conservative $\pm 1$-valued edge weight can be defined on a graft. 
Fixing a root determines a laminar family of sets of vertices according to their distance from the root. 
Seb\H{o}'s distance decomposition characterizes the structure of these families. 
We derive our constructive characterization
by proving that grafts with a CMJ satisfy a special property in terms of their distance decomposition.

We then use our constructive characterization to obtain a 
new polynomial-time recognition algorithm for grafts with a CMJ. 
Our algorithm runs in $O(n(m + n\log n))$ time and thus improves upon the previously known algorithm by Seb\H{o} and Tannier for sparse graphs. 
Our algorithm also finds a CMJ in $O(n(m + n\log n))$ time if one exists. 
Our algorithm has two bottlenecks: 
one is the procedure for computing an arbitrary minimum join for a given graft, 
and the other is the procedure for solving the single-source all-sink shortest path problem on graphs endowed with a conservative $\pm 1$-valued edge weight.  
We employ Gabow's result~\cite{10.1145/3183369} for the former and Gabow and Sankowski's~\cite{doi:10.1137/16M1106225} for the latter.  
All remaining steps run in  linear time. 
Therefore, improvements in these bottlenecks directly improve our algorithm.

The remainder of our paper is organized as follows. 
In Section~\ref{sec:definition}, we list basic terminology. 
In Section~\ref{sec:graft}, we explain fundamentals of joins and grafts. 
In Section~\ref{sec:sebo}, we introduce preliminary results due to Seb\H{o}~\cite{sebo1990} that are employed in this paper. 
Starting from Section~\ref{sec:rake}, 
we provide our new results. 
Section~\ref{sec:rake} introduces the new concept of rake grafts and derives their constructive characterization. 
Section~\ref{sec:cmj} derives our first main result, the constructive characterization of grafts with a CMJ. 
In the first half of Section~\ref{sec:alg}, 
we provide preliminary algorithmic results pertinent to our results, including the result by Seb\H{o} and Tannier~\cite{10.1007/3-540-45535-3_30, sebHo2004metric}.  
In the second half of Section~\ref{sec:alg},  
we derive our new algorithm.

\section{Basic Terminology}   \label{sec:definition}

We mostly follow Schrijver~\cite{schrijver2003} for standard notation. 
This section lists the exceptions and any nonstandard notation.  
We denote the symmetric difference of sets $A$ and $B$ by $A \Delta B$; that is, $A\Delta B = (A\setminus B) \cup (B\setminus A)$. 
We often treat an element $x$ as the singleton $\{x \}$.

We consider finite loopless multigraphs since loops do not play a substantial role in our discussions. 
All results in this paper also extend to finite multigraphs with loops. 
For two distinct vertices $x$ and $y$ in a graph, an edge between $x$ and $y$ can be denoted by $xy$. 
For convenience, we denote by $\{ xy :  y \in Y \}$, where $Y$ is a set of vertices, a set obtained by choosing exactly one edge between $x$ and $y$ for every $y\in Y$. 
For a graph $G$,  its vertex and edge sets are denoted by $V(G)$ and $E(G)$, respectively.  
We treat circuits and paths as graphs. 
That is, a circuit is a connected graph in which every vertex is of degree two. 
A path is a connected graph in which every vertex is of degree at most two and at least one vertex is of degree less than two. 
We often treat a graph $G$ as its vertex set $V(G)$.

Let $G$ be a graph,  and let $X, Y \subseteq V(G)$ and $F\subseteq E(G)$. 
The set of edges between $X$ and $Y$ is denoted by $E_G[X, Y]$. 
The set $E_G[X, V(G) \setminus X]$ is denoted by $\parcut{G}{X}$. 
The set of edges with both ends in $X$ is denoted by $E_G[X]$. 
We denote the set of edges in $F$ that span $X$ by $F[X]$. 
A neighbor of $X$ is a vertex in $V(G)\setminus X$ adjacent to a vertex in $X$. 
The set of neighbors of $X$ is denoted by $\parNei{G}{X}$.

Let $G$ be a graph,  and let $X \subseteq V(G)$. 
The subgraph of $G$ induced by $X$ is denoted by $G[X]$. 
The graph $G[V(G)\setminus X]$ is denoted by $G - X$. 
The graph obtained from $G$ by contracting $X$  is denoted by $G/X$. 
That is, 
$G/X$ is the graph obtained by replacing the vertices in $X$ by a single vertex $x$, 
removing all edges in $E_G[X]$, 
and replacing each edge in $E_G[X, y]$ by a new edge $xy$ for all $y\in V(G)\setminus X$. 
We denote the new vertex $x$ of $G/X$ by $[X]$.  
Let $\mathcal{X} = \{ X_ 1, \ldots, X_k \}$, where $k\ge 1$, be a family of disjoint subsets of $V(G)$. 
We denote the graph $G/X_1/X_2/\cdots/ X_k$ by  $G/\mathcal{X}$.  
The set $\{ [X_i] : i \in \{1,\ldots k\} \}$  of vertices in $G/\mathcal{X}$ is denoted by $\llbracket \mathcal{X} \rrbracket$. 

The sum of two graphs $G$ and $H$ is denoted by $G + H$. 
Let $H$ be a supergraph of $G$, and let $F \subseteq E(H)$. 
The graph obtained from $G$ by deleting $F$ is denoted by $G -F$; 
that is, $V(G-F) = V(G)$ and $E(G - F) = E(G)\setminus F$. 
The graph obtained from $G$ by adding $F$  is denoted by $G+F$. 
 Regarding operations on graphs, 
we identify the items of the new graph, such as edges and subgraphs, 
with the corresponding items of the old graph.

Let $G$ be a graph. 
We say that a set $F\subseteq E(G)$ of edges {\em covers} a vertex $v\in V(G)$  if $F$ has an edge incident to $v$. 
A set $M \subseteq E(G)$ of edges is a {\em matching} if every two distinct edges in $M$ are disjoint. 
A matching of $G$ is said to be {\em perfect} (resp. {\em  near-perfect}) if it covers all vertices (resp. all vertices except one)  in $G$. 
A graph $H$ is said to be {\em factor-critical} if $H-v$ has a perfect matching for every $v\in V(H)$.

\section{Grafts and Joins}  \label{sec:graft} 

In this section, we present basic definitions and properties of grafts and joins. 

\begin{definition} 
Let $G$ be a graph, and let $T\subseteq V(G)$. 
A pair $(G, T)$  is a {\em graft} 
if every connected component $C$ of $G$ has an even number of vertices from $T$.  
A set $F\subseteq E(G)$ is a {\em join} of the pair $(G, T)$ 
if $|\parcut{G}{v} \cap F|$ is odd for every $v\in T$ and  even for every $v\notin T$. 
\end{definition} 

A join of a graft $(G, T)$ is also known as a $T$-join of $G$. 
The handshaking lemma easily implies the following fact. 
(See, for example, Section 29.1 in Schrijver~\cite{schrijver2003}.)

\begin{fact}   \label{fact:graft}

For a graph $G$ and $T\subseteq V(G)$, 
the pair $(G, T)$ has a join if and only if it is a graft. 
\end{fact}

Under Fact~\ref{fact:graft},  
for a graft $(G, T)$, 
we call a join of $(G, T)$ with the minimum number of edges a {\em minimum} join 
and denote the number of edges in a minimum join of $(G, T)$ by $\nu(G, T)$.  
We say that a minimum join $F$ is {\em connected} if the subgraph of $G$ determined by $F$ is connected. 
If $F = \emptyset$, then we define that $F$ is {\em not} connected.

For a graft $(G, T)$, we often refer to the objects of $G$, such as vertices, edges,  and subgraphs, as the objects of $(G, T)$. 
We say that a graft $(G, T)$ is {\em connected} if $G$ is connected.

\begin{definition} 
Let $(G, T)$ be a graft, and let $F\subseteq E(G)$. 
We define the {\em $F$-weight} of an edge $e\in E(G)$ as $1$ and $-1$ according to the cases $e\notin F$ and $e\in F$ 
and denote it by $w_F(e)$. 
For a set $M$ of edges,  we define the {\em $F$-weight} of $M$ as $w_F(M) := \sum_{e\in M} w_F(e)$. 
For a subgraph $H$ of $G$, we  define its {\em $F$-weight} as $w_F(H):= w_F(E(H))$. 
\end{definition}

The next lemma characterizes minimum joins and is used in deriving our results.

\begin{lemma}[see Seb\H{o}~\cite{sebo1990}] \label{lem:circuit} 
Let $(G, T)$ be a graft, and let $F$ be a join of $(G, T)$. 
Then, $F$ is a minimum join 
if and only if $w_F(C) \ge 0$ holds for every circuit $C$ of $(G, T)$. 
\end{lemma}

Fundamental operations on grafts used in this paper are listed as follows.  
Let $(G, T)$ be a graft, and let $F$ be a minimum join of $(G, T)$. 
For $X \subseteq V(G)$,  
let $T_{F, X} \subseteq X$ be defined as follows: 
 A vertex $v\in X$ is an element of $T_{F, X}$ if and only if $| \parcut{G}{v} \cap F[X] |$ is odd. 
It is easily observed that $(G[X], T_{F, X})$ is a graft, and we denote this graft by $(G, T)_F[X]$. 
If $F\cap \parcut{G}{X} = \emptyset$, we sometimes denote $(G, T)_F[X]$ by $(G, T)[X]$, omitting the subscript $F$. 

We define $(G, T)/X$ as the graft obtained by contracting the set $X$ into a single vertex. 
More precisely,  
$(G, T)/X$ is defined as the graft $(G/X, T/X)$, where $T/X \subseteq V(G/X)$ is the set $( T \setminus X ) \cup  \{ [X] \}$ 
if $|X \cap T|$ is odd, and   the set $T \setminus X$ otherwise.  
For a family of disjoint subsets $\mathcal{X}$ of $V(G)$, $(G, T)/\mathcal{X}$ is defined in a similar way.

\section{Distances in Grafts and Seb\H{o}'s Theorems}  \label{sec:sebo} 
\subsection{Definition and Fundamental Properties of Distances in Grafts}

\begin{definition} 
Let $(G, T)$ be a graft, and let $F\subseteq E(G)$. 
For $x, y\in V(G)$, 
the minimum of $w_F(P)$ where $P$ ranges over all paths between $x$ and $y$ in $G$  is called the {\em $F$-distance} between $x$ and $y$, 
and is denoted by $\distgtf{G}{T}{F}{x}{y}$.  
A path between $x$ and $y$ is said to be {\em $F$-shortest}  if its $F$-weight is $\distgtf{G}{T}{F}{x}{y}$. 
If $G$ has no path between $x$ and $y$, then $\distgtf{G}{T}{F}{x}{y}$ is defined to be $\infty$. 
\end{definition}

The next three results introduce fundamental properties of distances in a graft 
that are sometimes used implicitly in this paper.

\begin{fact}[Seb\H{o}~\cite{sebo1990}]  \label{fact:distcanonical} 
Let $(G, T)$ be a graft,  let $x, y\in V(G)$, and let $F_1$ and $F_2$ be two minimum joins of $(G, T)$. 
Then, $\distgtf{G}{T}{F_1}{x}{y} = \distgtf{G}{T}{F_2}{x}{y}$. 
\end{fact}

Under Fact~\ref{fact:distcanonical}, 
we abbreviate $\distgtf{G}{T}{F}{x}{y}$ as $\distgt{G}{T}{x}{y}$, omitting the parameter $F$.

\begin{proposition}[Seb\H{o}~\cite{sebo1990}] \label{prop:neidist} 
Let $(G, T)$ be a connected graft, let $r\in V(G)$, and let $x, y\in V(G)$ be adjacent vertices. 
Then $|\distgt{G}{T}{r}{x} - \distgt{G}{T}{r}{y} | \le 1$.
\end{proposition}

\begin{proposition}[Seb\H{o}~\cite{sebo1990}] \label{prop:distalt} 
Let $(G, T)$ be a connected graft,  and let $r, r' \in V(G)$ be such that $r\neq r'$. 
Then, for every $x\in V(G)$, 
$\nu(G, T\Delta \{r, r'\}) = \nu(G, T) - \distgt{G}{T}{r}{r'}$, and 
$\distgt{G}{T\Delta \{r, r'\} }{r}{x} = \distgt{G}{T}{r}{x} - \distgt{G}{T}{r}{r'}$. 
\end{proposition}

\subsection{Strong Combs and Factor-Critical Grafts} 

We define two classes of grafts, strong combs and factor-critical grafts, 
that serve as the fundamental building blocks of a graft in terms of distance.

\begin{definition} 
Let $(G, T)$ be a graft.   
The graft $(G, T)$ is called a {\em strong comb}  with respect to $r \in V(G)$  with {\em tooth set} $B\subseteq V(G)$
if $B$ is a stable set with $\parNei{G}{B} = V(G)\setminus B$,  
$r$ is in $V(G)\setminus B$, 
and $\distgt{G}{T}{r}{x} = -1$ for every $x\in B$, whereas $\distgt{G}{T}{r}{x} = 0$ for every $x\in V(G)\setminus B$.  
\end{definition}

\begin{fact}[Seb\H{o}~\cite{sebo1990}]  \label{fact:comb} 
Let $(G, T)$ be a strong comb with tooth set $B$, and let $F$ be a minimum join of $(G, T)$. 
Then, $|\parcut{G}{v} \cap F| = 1$ for every $v\in B$, and $E_G[V(G)\setminus B] \cap F = \emptyset$. 
\end{fact}

\begin{definition} 
A graft $(G, T)$ is said to be {\em factor-critical}  with root $r\in V(G)$ 
if $G$ is factor-critical and $T = V(G) \setminus \{r\}$.  
\end{definition}

It is easily observed that  if a graft $(G, T)$ is factor-critical with root $r\in V(G)$, 
then every minimum join of $(G, T)$ is a near-perfect matching that covers all vertices of $G$ except $r$. 
This further implies that $\distgt{G}{T}{r}{x} = 0$ holds for every $x\in V(G)$, 
owing to a classical fact in matching theory~\cite{lp1986}.

\subsection{Seb\H{o}'s Distance Decomposition}  \label{sec:dist:sebo}

In this section, we present the theorems due to Seb\H{o}~\cite{sebo1990} on distances in grafts determined by a minimum join. 
Throughout this section, let $(G, T)$ be a connected graft, let $r\in V(G)$, 
and let $F$ be a minimum join, unless stated otherwise.  
Keep Proposition~\ref{prop:neidist} in mind to understand the following concepts and results.

\begin{definition} 
We denote the set $\{ \distgt{G}{T}{r}{x}  : x\in V(G) \}$ by $\interval{G}{T}{r}$.  
Let $i\in \interval{G}{T}{r}$. 
We define the set $\levelgtr{G}{T}{r}{i}$ as $\{ x\in V(G) : \distgt{G}{T}{r}{x} = i \}$,  
and the set $\laylegtr{G}{T}{r}{i}$ as $\{ x \in V(G) : \distgt{G}{T}{r}{x} \le i \}$. 
We denote by $\laycompgtr{G}{T}{r}{i}$ the set of connected components of $G[\laylegtr{G}{T}{r}{i}]$. 
We denote by $\qcompgtr{G}{T}{r}{i}$ the set of connected components of $G[\laylegtr{G}{T}{r}{i}] - E_G[\levelgtr{G}{T}{r}{i}]$. 
We denote the set $\bigcup_{i \in \interval{G}{T}{r}} \laycompgtr{G}{T}{r}{i}$ by $\laycompall{G}{T}{r}$, 
and the set $\bigcup_{i \in \interval{G}{T}{r}} \qcompgtr{G}{T}{r}{i} $ by $\qcompall{G}{T}{r}$. 
The set $\{ K \in \laycompall{G}{T}{r} : r\notin V(K) \}$ is denoted by $\noncapall{G}{T}{r}$, 
and the set $\{ K \in \qcompall{G}{T}{r} : r\notin V(K) \}$ is denoted by $\qnoncapall{G}{T}{r}$. 
\end{definition} 

By definition, $\{ \levelgtr{G}{T}{r}{i} :  i \in \interval{G}{T}{r} \}$ is a partition of $V(G)$, 
and the members of $\{ \laylegtr{G}{T}{r}{i} : i \in \interval{G}{T}{r} \}$ form an increasing family. 
By Proposition~\ref{prop:neidist}, every edge of $G$ 
 joins two vertices that lie in the same member of the partition or in two consecutive ones. 
Also, $\interval{G}{T}{r}$ is a set of consecutive integers. 
Note $r \in \levelgtr{G}{T}{r}{0}$. 
Consequently,  for every $i\in \interval{G}{T}{r}$, we have $r \notin \laylegtr{G}{T}{r}{i}$ if $i < 0$, 
and $r \in \laylegtr{G}{T}{r}{i}$ if $i \ge 0$.

\begin{definition} 
We denote the member $K \in \laycompgtr{G}{T}{r}{0}$ with $r\in V(K)$ by $\initialgtr{G}{T}{r}$ 
and the member $K \in \qcompgtr{G}{T}{r}{0}$ with $r\in V(K)$ by $\qinitialgtr{G}{T}{r}$.  
The set $\initialgtr{G}{T}{r} \cap \levelgtr{G}{T}{r}{0}$ is denoted by $\agtr{G}{T}{r}$, 
and the set $\initialgtr{G}{T}{r} \setminus \levelgtr{G}{T}{r}{0}$ is denoted by $\dgtr{G}{T}{r}$. 
\end{definition}

\begin{definition}  
Let $i\in \interval{G}{T}{r}$, and let $K \in \laycompgtr{G}{T}{r}{i} \cup \qcompgtr{G}{T}{r}{i}$. 
We denote the set $V(K)\cap \levelgtr{G}{T}{r}{i}$ by $\ak{G}{T}{K}$ 
and the set $V(K)\setminus \levelgtr{G}{T}{r}{i}$ by $\dk{G}{T}{K}$. 
We denote the set of connected components of $K[\dk{G}{T}{K}]$ by $\dkconn{G}{T}{K}$. 
For $K \in \laycompgtr{G}{T}{r}{i}$, 
we denote the set of connected components of $K - E_G[\levelgtr{G}{T}{r}{i}]$ 
by $\qkconn{G}{T}{K}$. 
\end{definition} 

The graph $K - E_G[\levelgtr{G}{T}{r}{i}]$ may be connected. 
Note that $\bigcup \{ \qkconn{G}{T}{K} : K \in \laycompgtr{G}{T}{r}{i} \} = \qcompgtr{G}{T}{r}{i}$ for every $i\in \interval{G}{T}{r}$.

In the remainder of this section, we present  part of the main results of Seb\H{o} that are used in this paper, 
including the definition of $F$-beams and -roots. 
Theorem~\ref{thm:sebo:beam} is the core of Seb\H{o}'s results. 
Theorems~\ref{thm:sebo:path} and \ref{thm:sebo:bone} show the recursive structure of the distances in a graft.

\begin{theorem}[Seb\H{o}~\cite{sebo1990}] \label{thm:sebo:beam} 
Let  $K \in \laycompall{G}{T}{r} \cup \qcompall{G}{T}{r}$.  
The following properties hold. 
\begin{rmenum} 
\item \label{item:sebo:beam:cap} If $r \in V(K)$ holds, then $\parcut{G}{K} \cap F = \emptyset$. 
\item \label{item:sebo:beam:noncap} If $r \notin V(K)$ holds, then $| \parcut{G}{K} \cap F |= 1$.  
\end{rmenum} 
\end{theorem}

\begin{definition} 
Under Theorem~\ref{thm:sebo:beam}, 
for $K\in \noncapall{G}{T}{r} \cup \qnoncapall{G}{T}{r}$, 
we call the unique edge in $\parcut{G}{K} \cap F$ the {\em $F$-beam} of $K$. 
We call the end of the $F$-beam of $K$ that is in $V(K)$  the {\em $F$-root} of $K$. 
\end{definition}

\begin{theorem} [Seb\H{o}~\cite{sebo1990}]   \label{thm:sebo:path} 
Let $K \in \noncapall{G}{T}{r}\cup \qnoncapall{G}{T}{r} \cup \{ \initialgtr{G}{T}{r}, \qinitialgtr{G}{T}{r}\}$.  
Let $r_K$ be the $F$-root of $K$ if $K \in \noncapall{G}{T}{r}\cup \qnoncapall{G}{T}{r}$, and $r_K = r$ if $K \in \{ \initialgtr{G}{T}{r}, \qinitialgtr{G}{T}{r}\}$. 
Then, the following properties hold: 
\begin{rmenum} 
\item $F[K]$ is a minimum join of $(G, T)_F[K]$. 
\item  \label{item:sebo:path:project} 
$\distgtf{G}{T}{F}{r}{x} = \distgtf{G}{T}{F}{r}{r_K} + \distpf{ (G, T)_F[K] }{F[K]}{r_K}{x}$ for every $x\in V(K)$. 
\end{rmenum} 
\end{theorem}

By Proposition~\ref{prop:neidist}, 
$\distgtf{G}{T}{F}{r}{x} \le \distgtf{G}{T}{F}{r}{r_K}$
holds in Theorem~\ref{thm:sebo:path} \ref{item:sebo:path:project}. 
 Hence, $\distpf{ (G, T)_F[K] }{F[K]}{r_K}{x} \le 0$. 
In particular, 
$\distpf{ (G, T)_F[K] }{F[K]}{r_K}{x} = 0$ 
for every $x\in \ak{G}{T}{K}$, 
whereas $\distpf{ (G, T)_F[K] }{F[K]}{r_K}{x} < 0$ for every $x\in \dk{G}{T}{K}$.

\begin{theorem}[Seb\H{o}~\cite{sebo1990}]  \label{thm:sebo:bone} 
Let $K \in \noncapall{G}{T}{r} \cup \qnoncapall{G}{T}{r} \cup \{\initialgtr{G}{T}{r}, \qinitialgtr{G}{T}{r} \}$, 
and let $r_K$ be the $F$-root of $K$ for $K \in \noncapall{G}{T}{r} \cup \qnoncapall{G}{T}{r}$,  
and $r$ for $K \in \{ \initialgtr{G}{T}{r}, \qinitialgtr{G}{T}{r} \}$. 
Then, the following properties hold: 
\begin{rmenum} 
\item \label{item:sebo:q} 
If $K \in \noncapall{G}{T}{r} \cup \{ \initialgtr{G}{T}{r}\}$, then 
  $(G, T)_F[K] / \qkconn{G}{T}{K}$ is a factor-critical graft whose root is $[Q_0]$, where $Q_0$ is the member of $\qkconn{G}{T}{K}$ with $r_K\in V(Q_0)$. 
Furthermore, $F[\ak{G}{T}{K}]$ is a minimum join of $(G, T)_F[K] / \qkconn{G}{T}{K}$. 
\item \label{item:sebo:d} 
If  $K \in \qnoncapall{G}{T}{r} \cup \{ \qinitialgtr{G}{T}{r} \}$,  
then $(G, T)_F[K]/\dkconn{G}{T}{K}$ is a strong comb with respect to $r_K$ with the tooth set $\llbracket \dkconn{G}{T}{K}  \rrbracket$. 
Furthermore, $F[K] \setminus E_G[\dk{G}{T}{K}]$ is a minimum join of $(G, T)_F[K]/\dkconn{G}{T}{K}$. 
\end{rmenum} 
\end{theorem}

We call the structure of grafts described by Theorems~\ref{thm:sebo:beam}, \ref{thm:sebo:path}, and \ref{thm:sebo:bone} 
the {\em distance decomposition}.

\section{Rakes and Their Constructive Characterization}  \label{sec:rake}

This section introduces a new concept called rakes and presents their constructive characterization. 
Rakes are strong combs with a connected minimum join covering the root, 
and serve as fundamental building blocks of grafts with a connected minimum join.

\begin{definition} 
We call a graft $(G, T)$ a {\em rake}  with {\em tooth set} $B$  and {\em head} $r$ 
if $B$ is a subset of $T$ stable in $G$, $\parNei{G}{B} = V(G)\setminus B$, 
$r$ is a vertex in $V(G)\setminus B$ and is adjacent to every vertex in $B$, and 
$T = B \cup \{r\}$ when $|B|$ is odd, whereas $T = B$ when $|B|$ is even. 
\end{definition}

\begin{definition} 
Let $G$ be a graph, let $r \in V(G)$, and let $B \subseteq V(G)$ be a nonempty set of vertices. 
We call a set of edges of the form $\{ rb : b\in B\}$ a {\em $(B, r)$-star set}. 
That is, a set $F\subseteq E_G[r, B]$ is a $(B, r)$-star set if $|F\cap E_G[r, b]| = 1$ for every $b\in B$. 
\end{definition}

Observations~\ref{obs:rake2stcomb} and \ref{obs:connjoin2rake} are easily derived and might be of help in understanding the role of rakes in our results.

\begin{observation}  \label{obs:rake2stcomb} 
Let $(G, T)$ be a rake with  head $r$ and tooth set $B$.  
Then, $(G, T)$ has a $(B, r)$-star set as a connected minimum join and is a strong comb with respect to $r$. 
\end{observation} 
\begin{proof}  Let $F\subseteq E(G)$ be a $(B, r)$-star set. 
Then $F$ is clearly a join of $(G, T)$. Additionally, since $B$ is a stable subset of $T$, we have $\nu(G, T)  \ge |B| = |F|$.  
Hence, $F$ is a minimum join, which is connected. 
It is easily observed that  $\distgtf{G}{T}{F}{r}{x} = -1$ for every $x\in B$,  
and $\distgtf{G}{T}{F}{r}{x} = 0$ for every $x \in V(G)\setminus B$. 
Therefore, $(G, T)$ is a strong comb with respect to $r$. 
\end{proof}

\begin{observation}  \label{obs:connjoin2rake} 
Let $(G, T)$ be a strong comb with respect to $r$ with tooth set $B$.  
Then,  $(G, T)$ has a connected minimum join covering $r$ if and only if it is a rake with head $r$ and tooth set $B$. 
\end{observation} 
\begin{proof} 
If $(G, T)$ has a connected minimum join $F$,  
then Fact~\ref{fact:comb} implies that $F$ is a $(B, v)$-star set for some vertex $v \in V(G)\setminus B$.  
As $F$ is assumed to cover $r$, this implies $v = r$. 
Hence, $(G, T)$ is a rake with head $r$ and tooth set $B$.  
Conversely, if $(G, T)$ is a rake with head $r$ and tooth set $B$, 
then a $(B, r)$-star set is clearly a connected minimum join of $(G, T)$.  
\end{proof}

\begin{definition} 
Define $\mathcal{R}(r, B)$ to be the smallest family of grafts with a vertex $r$ and a nonempty set of vertices $B$ with $r\notin B$ satisfying the following: 
\begin{rmenum} 
\item \label{item:rake:star} Let $H$ be a graph with $V(H) = \{r\} \dot\cup B$ and $E(H) = \{ rx : x\in B \}$. 
 Let $T = V(H)$ if $|V(H)|$ is even, and $T = V(H)\setminus \{r\}$  otherwise. 
Then, the graft $(H, T)$ is a member of $\mathcal{R}(r, B)$. 
\item \label{item:rake:vadd} Let $(G, T) \in \mathcal{R}(r, B)$, and let $x\not\in V(G)$. 
Let $\hat{G}$ be a graph such that $V(\hat{G}) = V(G) \cup \{x \}$ and $E(\hat{G}) = E(G) \cup F$, where 
$F$ is a nonempty set of edges between $\{x\}$ and $B$. 
Then, the graft $(\hat{G}, T)$ is a member of $\mathcal{R}(r, B)$. 
\item \label{item:rake:eadd} Let $(G, T) \in \mathcal{R}(r, B)$. 
Let $\hat{G}$ be a graph with $V(\hat{G}) = V(G)$ and $E(\hat{G}) = E(G)\cup F$, 
where $F$ is a set of edges between two vertices in $V(G)\setminus B$ or between $B$ and $V(G)\setminus B$. 
Then the graft $(\hat{G}, T)$ is a member of $\mathcal{R}(r, B)$. 
\end{rmenum} 
\end{definition}

We provide a constructive characterization of rakes as follows.

\begin{proposition}  \label{prop:rakechar} 
For a graph $G$ and a set $T\subseteq V(G)$, 
the pair $(G, T)$ is a member of $\mathcal{R}(r, B)$ 
if and only if $(G, T)$ is a rake with  tooth set $B$  and head $r$.  
\end{proposition} 
\begin{proof} 
We first prove the necessity part. 
Let $(G, T)$ be a rake with  tooth set $B$  and head $r$. 
By the definition of rakes, 
$E(G)$ contains a $(B, r)$-star set $F$; further, $T = B$ if $|B|$ is even, and  $T = B \cup \{r\}$ otherwise.  
Let $(H, T)$ be a subgraft of $G$ such that $H = (\{r\}\cup B, F)$. 
If $(G, T) = (H, T)$,  then $(G, T)$ is clearly a member of $\mathcal{R}(r,B)$. 
Otherwise, $(G,T)$ can be obtained from $(H,T)$  
by successively applying the constructions in \ref{item:rake:vadd} and \ref{item:rake:eadd}. 
Thus, $(G, T) \in \mathcal{R}(r, B)$ holds. 

Next, we prove the sufficiency part. 
The graft $(H,T)$ defined in \ref{item:rake:star} is clearly a rake with tooth set $B$ and head $r$.  
If a graft $(G,T)$ is a member of $\mathcal{R}(r, B)$, then the graft $(\hat{G},T)$ obtained from $(G,T)$
as described in \ref{item:rake:vadd} is also a rake with tooth set $B$ and head $r$.
The same argument applies to the graft obtained in \ref{item:rake:eadd}.
\end{proof}

\section{Characterization of Grafts with Connected Minimum Joins}  \label{sec:cmj}

\subsection{Necessary Condition for Grafts with a Connected Minimum Join}

In this section, we present Lemma~\ref{lem:connjoin2char}, 
which establishes properties of grafts with a connected minimum join, to be used in deriving their characterization in Theorem~\ref{thm:char}. 
Lemma~\ref{lem:connjoin2char} proves that  
if a graft $(G, T)$ has a connected minimum join covering $r\in V(G)$, then $\initialgtr{G}{T}{r}$ is a recursive combination of rakes, 
whereas $G - \initialgtr{G}{T}{r}$ serves as a ``tail'' 
that is disjoint from any minimum join.

\begin{lemma}  \label{lem:connjoin2char} 
Let $(G, T)$ be a connected graft with a connected minimum join $F$, 
and let $r\in V(G)$ be a vertex covered by $F$.  
Then, the following properties hold. 
\begin{rmenum} 
\item \label{item:nont} 
$F\subseteq E(\initialgtr{G}{T}{r})$ holds. Accordingly, $T \subseteq V(\initialgtr{G}{T}{r})$ holds. 
\item \label{item:nontri} 
$|V(\initialgtr{G}{T}{r})| > 1$.  
\item \label{item:tailnei} 
$\laycompgtr{G}{T}{r}{0} = \{ \initialgtr{G}{T}{r} \}$.  
Also, $\parNei{G}{G - \initialgtr{G}{T}{r}} \subseteq \ak{G}{T}{\initialgtr{G}{T}{r}}$ holds. 
\item \label{item:nonq} \label{item:init} 
$| \qkconn{G}{T}{K} | = 1$ for every $K \in \noncapall{G}{T}{r} \cup \{ \initialgtr{G}{T}{r} \}$. 
\item \label{item:rake} 
Let $K = \initialgtr{G}{T}{r}$, or let $K$ be a member of $\noncapall{G}{T}{r}$ with $\dk{G}{T}{K}\neq \emptyset$. 
Let $r_K$ be $r$ if $K = \initialgtr{G}{T}{r}$ and the $F$-root of $K$ otherwise. 
Then, the graft $(G, T)_F[K]/\dkconn{G}{T}{K}$ is a rake with head $r_K$ and tooth set $\llbracket \dkconn{G}{T}{K} \rrbracket$. 
\item  \label{item:joinproj} 
Furthermore, let $r_L$ be the $F$-root of $L$ for every $L \in \dkconn{G}{T}{K}$; 
$F[K] \setminus E_G[\dk{G}{T}{K}]$ equals $\{ r_Kr_L: L \in \dkconn{G}{T}{K} \}$, 
which is a connected minimum join of the rake $(G, T)_F[K]/\dkconn{G}{T}{K}$. 
\end{rmenum} 
\end{lemma} 
\begin{proof}According to Theorem~\ref{thm:sebo:beam} \ref{item:sebo:beam:cap}, no edge in $\parcut{G}{\initialgtr{G}{T}{r}}$ is in $F$. 
Because $F$ covers $r$,  this implies $F \subseteq E(\initialgtr{G}{T}{r})$.  
Consequently, we obtain $T \subseteq V(\initialgtr{G}{T}{r})$.  
It also follows that $V(\initialgtr{G}{T}{r}) \neq \{r\}$. 
The statements \ref{item:nont} and \ref{item:nontri}  are proved.  
Furthermore, by Theorem~\ref{thm:sebo:beam} \ref{item:sebo:beam:noncap}, $F \subseteq E(\initialgtr{G}{T}{r})$ also implies $\laycompgtr{G}{T}{r}{0} = \{ \initialgtr{G}{T}{r} \}$. Thus, $V(G - \initialgtr{G}{T}{r}) \subseteq \{ x\in V(G): \distgt{G}{T}{r}{x} > 0\}$. 
Therefore, Proposition~\ref{prop:neidist} implies $\parNei{G}{G - \initialgtr{G}{T}{r}} \subseteq \ak{G}{T}{\initialgtr{G}{T}{r}}$. 
The statement \ref{item:tailnei} is proved.

Suppose $| \qkconn{G}{T}{K} | >1$ for a member $K$ of $\noncapall{G}{T}{r} \cup \{ \initialgtr{G}{T}{r} \}$. 
Let $Q_0$ be the member of $\qkconn{G}{T}{K}$ that contains $r_K$, 
where $r_K$ is $r$ if $K = \initialgtr{G}{T}{r}$ and is the $F$-root of $K$ otherwise.  
Theorem~\ref{thm:sebo:bone} \ref{item:sebo:q} implies that no edge joining $Q_0$ and $K -V(Q_0)$ is in $F$ 
and that $\ak{G}{T}{K} \setminus V(Q_0)$ must contain an edge in $F$. 
However, Theorem~\ref{thm:sebo:beam} \ref{item:sebo:beam:noncap} 
or the assumption on $F$ implies that $F$ also has an edge in  either $\parcut{G}{r_K} \cap \parcut{G}{K}$ or $E(Q_0)$.   
This contradicts  $F$ being connected.  
Thus, \ref{item:nonq} is proved.

We next prove \ref{item:rake}  and \ref{item:joinproj}.  
By Theorem~\ref{thm:sebo:beam} and the assumption on $F$,        
$F[K]$ determines a connected subgraph in $K$ and covers $r_K$. 
Therefore, $F[K] \setminus E_G[\dk{G}{T}{K}]$ determines a connected subgraph in $K/\dkconn{G}{T}{K}$ and covers $r_K$. 
As \ref{item:nonq} implies $K \in \qnoncapall{G}{T}{r} \cup \{ \qinitialgtr{G}{T}{r} \}$, 
Theorem~\ref{thm:sebo:bone}~\ref{item:sebo:d} implies that 
$(G, T)_F[K]/\dkconn{G}{T}{K}$ is a strong comb with respect to $r_K$ 
and with tooth set $\llbracket \dkconn{G}{T}{K} \rrbracket$, 
for which $F[K]\setminus E_G[\dk{G}{T}{K}]$ is a minimum join; 
additionally, by Fact~\ref{fact:comb},  
this minimum join is equal to the set of $F$-beams of all members of $\dkconn{G}{T}{K}$. 
Therefore, we have $F[K]\setminus E_G[\dk{G}{T}{K}] = \{r_Kr_L: L \in \dkconn{G}{T}{K} \}$.  
This further implies that $(G, T)_F[K]/\dkconn{G}{T}{K}$ is a rake with head $r_K$ and tooth set $\llbracket \dkconn{G}{T}{K} \rrbracket$.  
Thus, \ref{item:rake}  and \ref{item:joinproj} are proved.   
This completes the proof of the lemma. 
\end{proof}

\subsection{Sufficient Condition for Grafts with a Connected Minimum Join}

In this section, we prove Lemmas~\ref{lem:biprimal2const} and \ref{lem:tail},
which together provide a sufficient condition for a graft to have a connected minimum join.

\begin{definition}  \label{def:gluesum} 
Let $(G_0, T_0)$ be a connected graft, and let $S \subseteq T_0$ be a stable set of size $k$, where $k\ge 1$. 
Let $e_s \in \parcut{G_0}{s}$ for every $s \in S$, and let $F := \{ e_s: s \in S \}$.  
Let $\{ (G_s, T_s): s \in S\}$ be a family of mutually disjoint grafts that are each disjoint from $(G_0, T_0)$. 
For each $s\in S$, let $A_s \subseteq V(G_s)$ and $r_s\in A_s$.  
For each $s \in S$,   
let $f_s: \parcut{G_0}{s} \to A_s$ be a mapping such that $r_s = f_s(e_s)$. 
Let a graph $G$ and  a set $T\subseteq V(G)$ be defined as follows: 
\begin{rmenum} 
\item $V(G) = ( V(G_0)\setminus S)  \cup \bigcup_{s\in S} V(G_s)$,  
\item $E(G) = ( E(G_0)\setminus  \parcut{G_0}{S})  \cup \bigcup_{s\in S} \{ xf_s(xs) : xs \in \parcut{G_0}{s}, x\in V(G_0)\setminus S \} 
\cup \bigcup_{s\in S} E(G_s)$,  and 
\item  $T = ( T_0 \setminus S) \cup \bigcup_{s\in S} (T_s \Delta \{r_s\})$.  
\end{rmenum} 
We call the graft $(G, T)$ a {\em gluing sum} of $(G_0, T_0)$ and $\{ (G_s, T_s) :s\in S\}$  
and denote it by $(G_0, T_0; S, F)\oplus \{ (G_s, T_s; A_s, r_s) : s\in S \}$. 
\end{definition} 

\begin{figure} \centering
\includegraphics[width=0.7\linewidth]{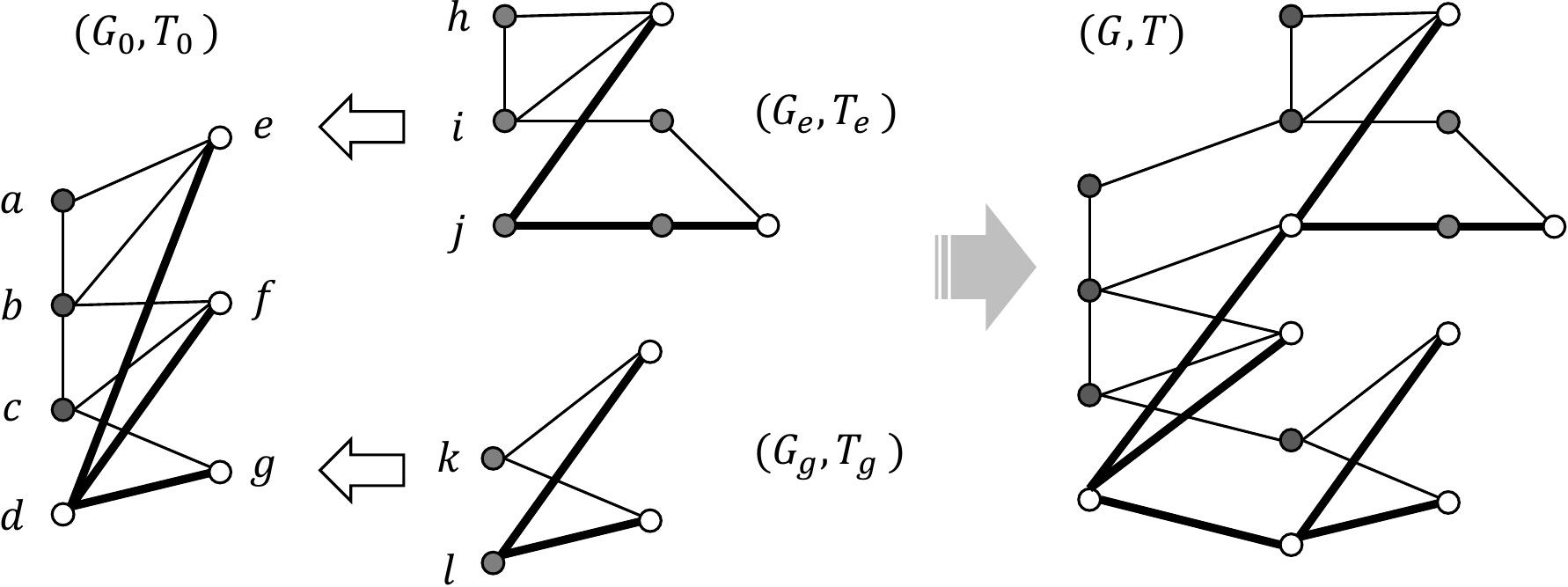}

  \vspace{14mm}

  \includegraphics[width=0.7\linewidth]{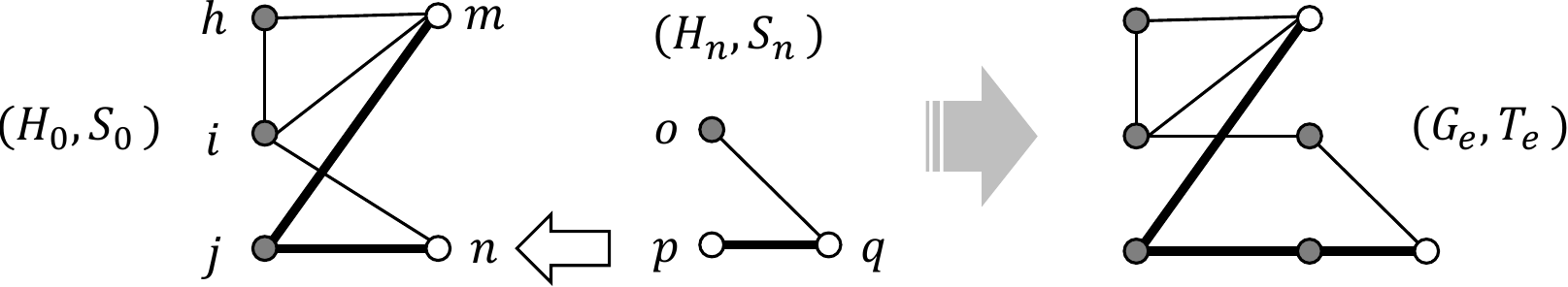} 
\vspace{5mm}
  \caption{Construction of a primal graft with a connected minimum join according to the gluing sum construction in Definition~\ref{def:primal}: 
    The graft $(G, T)$ is a gluing sum of $(G_0, T_0)$ and  $\{(G_e, T_e), (G_f, T_f), (G_g, T_g)\}$, 
     where $(G_0, T_0) \in \mathcal{R}(d, \{e, f, g\})$, $(G_e, T_e)\in \mathcal{P}(j, \{h, i, j\})$, and $(G_g, T_g) \in \mathcal{P}(l, \{ k, l\})$, whereas $(G_f, T_f)$ is trivial.   
     Thus $(G, T) \in \mathcal{P}(d, \{a, b, c, d\})$ holds.  
     Furthermore, the graft $(G_e, T_e)$ is a gluing sum of $(H_0, S_0; \{m, n\}, \{ jm, jn \})$ and $\{ (H_m, S_m), (H_n, S_n)\}$, 
     where $(H_0, S_0) \in \mathcal{R}(j, \{m, n\})$ and $(H_n, S_n)\in \mathcal{R}(p, \{q\})$, 
     whereas $(H_m, S_m)$ is trivial. 
     In each graft, the vertices in $T_0, T_e, T_f, T_g, S_0, S_m, S_n$, and $T$ are represented by white points. Bold lines indicate edges in a connected minimum join. }
\label{fig:synth}

\end{figure}

That is, $(G, T)$ in Definition~\ref{def:gluesum}  is a graft obtained as follows: 
$G$ is obtained from $G_0$ by replacing each $s \in S$ with $G_s$ so that 
the end $s$ of each edge in $\parcut{G_0}{s}$ is replaced by a vertex in $A_s$, 
and in particular the end $s$ of the designated edge $e_s$ is replaced by the designated vertex $r_s$; 
additionally, $T$ is obtained from $T_0$ by replacing each $s \in S$ with $T_s \Delta \{ r_s\}$. 
Note that a gluing sum is not necessarily uniquely determined 
due to the multiple choices for $f_s$.
 See also Figure~\ref{fig:synth}. 
Observation~\ref{obs:gluesum} is easily confirmed, and we use it in the sequel without explicit mention.

\begin{observation}   \label{obs:gluesum} 
Let $(G, T)$ be a graft, and assume that $(G_0, T_0; S, F)$  
 and $\{ (G_s, T_s; A_s, r_s) : s\in S \}$ are given so that their gluing sum is well-defined. 
For each $s \in S$, let $x_s \in V(G_0) \setminus S$ be such that $e_s = x_s s$. 
Then,  $(G, T)$ is a gluing sum $(G_0, T_0; S, F)\oplus \{ (G_s, T_s; A_s, r_s) : s\in S \}$ 
if and only if the following properties are satisfied:
\begin{rmenum} 
\item $\{ G_s : s\in S\}$ is a family of mutually disjoint and nonadjacent induced subgraphs of $G$,  and 
 contracting $G_s$ into a single vertex $s$ for every $s\in S$ results in $G_0$;   
\item for each $s\in S$, $x_sr_s \in \parcut{G}{G_s}$   and $\parNei{G}{G-G_s} \cap V(G_s) \subseteq A_s$ hold; and 
\item  $V(G_s)\cap T = T_s\Delta \{r_s\}$ and $T\cap ( V(G_0) \setminus S) = T_0\setminus S$.  
\end{rmenum} 
\end{observation}

\begin{definition} \label{def:primal} 
For every set $A$ and every $r \in A$, 
we simultaneously define the family $\mathcal{P}(r, A)$ of grafts $(G, T)$ with $A \subseteq V(G)$ 
as the smallest system of families satisfying the following conditions: 
\begin{rmenum} 
\item 
If a graft $(G, T)$ with $A \subseteq V(G)$ is a member of $\mathcal{R}(r, V(G)\setminus A)$, then it is a member of $\mathcal{P}(r, A)$.  
\item 
Let $(G, T)\in \mathcal{R}(r, B)$ be a graft with $V(G) = A \dot\cup B$. 
Let $\{ (H_b, T_b) : b\in B \}$ be a family of mutually disjoint grafts that are also disjoint from $(G, T)$, 
where  for each $b\in B$,  either $(H_b, T_b) \in \mathcal{P}(r_b, A_b)$ or $(H_b, T_b) = ( ( \{r_b\}, \emptyset), \emptyset)$ 
with $A_b = \{r_b\}$. 
Then, any gluing sum $(G, T; B, \{rb: b\in B\}) \oplus \{ (H_b, T_b; A_b, r_b) : b\in B\}$ is a member of $\mathcal{P}(r, A)$.  
\end{rmenum} 
\end{definition}

\begin{definition} 
A graft $(G, T)$ is said to be {\em primal} with respect to  $r\in V(G)$ 
if $\distgt{G}{T}{r}{x} \le 0$ for every $x\in V(G)$. 
\end{definition}

The next lemma provides a sufficient condition for primal grafts to have a connected minimum join. 
See also Figure~\ref{fig:synth}.

\begin{lemma}  \label{lem:biprimal2const} 
Let $(G, T) \in \mathcal{P}(r, A)$. 
Then, the following properties hold: 
\begin{rmenum} 
\item \label{item:primal} $(G, T)$ is a primal graft with respect to $r$ such that $A = \{ x \in V(G): \distgt{G}{T}{r}{x} = 0\}$. 
\item \label{item:join}  
If $V(G)\neq \{r \}$, then $(G, T)$ has a connected minimum join that covers $r$.  
\item \label{item:a2extreme} 
$\distgt{G}{T}{x}{y} \ge 0$ for every $x, y\in A$. 
\end{rmenum} 
\end{lemma} 
\begin{proof}We prove the lemma along the inductive definition on $\mathcal{P}(r, A)$. 
First, every member of $\mathcal{R}(r, V(G)\setminus A)$ clearly satisfies the claims \ref{item:primal}, \ref{item:join}, and \ref{item:a2extreme} because it is a rake with head $r$ and tooth set $V(G)\setminus A$, according to Proposition~\ref{prop:rakechar}.   
Also, note that \ref{item:primal}, \ref{item:join}, and \ref{item:a2extreme} hold for any trivial graft with $V(G) = \{r\}$.

Let $(\hat{G}, \hat{T})$ be a gluing sum $(G, T; B, \{rb: b\in B\} ) \oplus \{ (H_b, T_b; A_b, r_b) : b\in B\}$ 
such that $B = V(G) \setminus A$, 
where $(G, T)\in \mathcal{R}(r, B)$ and  $(H_b, T_b) \in \mathcal{P}(r_b, A_b)$ or $V(H_b) = \{r_b\}$ for each $b\in B$; 
note that $(G, T)$ is a rake with head $r$ and tooth set $B$, owing to Proposition~\ref{prop:rakechar}. 
Assume that the claims \ref{item:primal}, \ref{item:join}, and \ref{item:a2extreme} hold for $(G, T)$ and $(H_b, T_b)$'s. 
Let $F := \{ rr_b : b\in B\}$. 
For each $b \in B$, let $F_b$ be a connected minimum join of $(H_b, T_b)$ that covers $r_b$ if $(H_b, T_b)$ is in $\mathcal{P}(r_b, A_b)$; 
otherwise, let $F_b := \emptyset$.  
Let $\hat{F} := F \cup \bigcup_{b\in B} F_b$.

\begin{pclaim}  \label{claim:biprimal2const:nonneg} 
Let $C$ be a subgraph of $\hat{G}$ that is a circuit or a path between two vertices $x, y\in A$. 
Then, $w_{\hat{F}}(C) \ge 0$. 
\end{pclaim} 
\begin{proof}[Proof of Claim~\ref{claim:biprimal2const:nonneg}]
If $V(C)\subseteq A$ holds, then the claim obviously holds because $\hat{F}\cap E_{\hat{G}}[A] = \emptyset$. 
If $V(C) \subseteq V(H_b)$  for some $b\in B$, 
then the claim follows from Lemma~\ref{lem:circuit} or the induction hypothesis. 
In the following, we consider the case where neither of these conditions holds.  
From the hypothesis,  
for every $b\in B$ with $V(C)\cap V(H_b)\neq \emptyset$, every connected component of $C[H_b]$ is a path between two vertices in $A_b$ with nonnegative $F_b$-weight. 
Also,  
because $\parcut{\hat{G}}{H_b}$ contains only one edge in $\hat{F}$, 
we have $w_{\hat{F}}( E(C) \cap  \parcut{\hat{G}}{H_b}) \ge 0$. 
Furthermore, it is clear that $w_{\hat{F}}( E(C) \cap E_G[A] ) \ge 0$.  
It follows that $C$ has a nonnegative $\hat{F}$-weight. 
The claim is proved. 
\end{proof}

\begin{pclaim}  \label{claim:biprimal2const:join} 
$\hat{F}$ is a connected minimum join of $(\hat{G}, \hat{T})$ that covers $r$. 
\end{pclaim} 
\begin{proof}[Proof of Claim~\ref{claim:biprimal2const:join}]
It is clear that $\hat{F}$ is a connected join of $(\hat{G}, \hat{T})$.  
For every circuit $C$ in $\hat{G}$, 
Claim~\ref{claim:biprimal2const:nonneg} implies $w_{\hat{F}}(C) \ge 0$. 
Therefore, Lemma~\ref{lem:circuit} implies that $\hat{F}$ is a minimum join. 
Thus, the claim follows. 
\end{proof}

Claim~\ref{claim:biprimal2const:join} immediately proves \ref{item:join}. 
Claims~\ref{claim:biprimal2const:nonneg} and \ref{claim:biprimal2const:join} prove \ref{item:a2extreme}.

\begin{pclaim} \label{claim:biprimal2const:primal}  
Let $x\in V(\hat{G})$. 
\begin{rmenum} 
\item \label{item:primal:neg} 
If $x$ is a vertex in $\bigcup_{b\in B} V(H_b)$, then $\hat{G}$ has a path of negative $\hat{F}$-weight between $r$ and $x$. 
\item \label{item:primal:zero}  
If $x$ is a vertex in $A$, 
then $\hat{G}$ has a path of zero $\hat{F}$-weight between $r$ and $x$.  
\end{rmenum} 
\end{pclaim} 
\begin{proof}[Proof of Claim~\ref{claim:biprimal2const:primal}] 
Let $b\in B$ and $x \in V(H_b)$. 
From the induction hypothesis, $(H_b, T_b)$ is a primal graft with respect to $r_b$ and thus has a path $P$ between $x$ and $r_b$ with $w_{F_b}(P) \le 0$. 
Then, $P + r_br$ is a path between $x$ and $r$ with negative $\hat{F}$-weight. 
This proves \ref{item:primal:neg}. 

Next, let $x\in A$.  If $x = r$, then \ref{item:primal:zero} trivially holds. Assume $x\neq r$. 
As $(G, T)$ is a rake, there exists $b\in B$ with $xb\in E(G)$; 
let $y\in A_b$ be the vertex of $\hat{G}$ such that $xb$ corresponds to $xy$ in $\hat{G}$.    
From $y, r_b\in A_b$, the induction hypothesis implies that 
$H_b$ has a path $Q$ between $y$ and $r_b$ whose $F_b$-weight is $0$. 
Thus, $Q + xy + r_br$ is a path in $\hat{G}$ between $x$ and $r$ whose $\hat{F}$-weight is $0$.   
This proves \ref{item:primal:zero}. 
\end{proof}

Claims~\ref{claim:biprimal2const:join} and \ref{claim:biprimal2const:primal} imply that $(\hat{G}, \hat{T})$ is a primal graft with respect to $r$. 
Claim~\ref{claim:biprimal2const:nonneg} further implies $A \subseteq \{ x \in V(\hat{G}): \distgt{\hat{G}}{\hat{T}}{r}{x} = 0\}$. 
Additionally, Claim~\ref{claim:biprimal2const:primal} \ref{item:primal:neg} implies $A \supseteq \{ x \in V(\hat{G}) : \distgt{\hat{G}}{\hat{T}}{r}{x} = 0\}$. 
Accordingly, $A = \{ x \in V(\hat{G}) : \distgt{\hat{G}}{\hat{T}}{r}{x} = 0\}$, and thus \ref{item:primal} is proved. 
The lemma is proved. 
\end{proof}

In contrast to Lemma~\ref{lem:biprimal2const}, 
the next lemma provides a sufficient condition for nonprimal grafts to have a connected minimum join.

\begin{lemma}  \label{lem:tail} 
Let $(G, T) \in \mathcal{P}(r, A)$, and let $(H, \emptyset)$ be a graft with $V(G)\cap V(H) = \emptyset$. 
Then, for any set $S$ of edges between $A$ and $V(H)$, 
the pair $(G + H + S, T)$ is a graft with a connected minimum join that covers $r$. 
\end{lemma}  
\begin{proof}Under Lemma~\ref{lem:biprimal2const}, 
let $F$ be a connected minimum join of $(G, T)$ that covers $r$. 
It is easily confirmed that $F$ is a join of $(G + H + S, T)$. 
We prove its minimality in the following. 
Let $C$ be a circuit of $(G + H + S, T)$.  
If $V(C) \subseteq V(G)$ or $V(C) \subseteq V(H)$ holds, then $w_F(C) \ge 0$ immediately follows from Lemma~\ref{lem:circuit}. 
Assume that $C$ is not a subgraph of either $G$ or $H$. 
Then, 
every connected component of $C[G]$ is a path between two vertices in $A$. 
Lemma~\ref{lem:biprimal2const} implies that these paths have nonnegative $F$-weights. 
Thus, $w_F(E(C)\cap E(G)) \ge 0$. 
Also, it is obvious that $w_F(E(C) \setminus E(G)) > 0$. 
Hence, 
we have $w_F(C) \ge 0$.  
Therefore, Lemma~\ref{lem:circuit} implies that $F$ is a minimum join of $(G + H + S, T)$.   
This proves the lemma. 
\end{proof}

\subsection{Characterization} 

Theorem~\ref{thm:char} gives a constructive characterization of grafts with a connected minimum join.

\begin{theorem}  \label{thm:char} 
The following two properties are equivalent for a connected graft $(G, T)$:  
\begin{rmenum} 
\item \label{item:char:join} The graft $(G, T)$ has a connected minimum join that covers $r\in V(G)$. 
\item \label{item:char:p} The graft $(G, T)$ is a member of $\mathcal{P}(r, A)$ for a set $A \subseteq V(G)$ and a vertex $r\in A$, 
or $(G, T)$ is obtained from a graft $(G', T) \in \mathcal{P}(r, A)$ 
and a graft $(H, \emptyset)$ with $V(H)\cap V(G') = \emptyset$ by adding some edges between $V(H)$ and $A$. 
\end{rmenum} 
\end{theorem} 
\begin{proof} Assume that \ref{item:char:join} holds. Let $F$ be a connected minimum join of $(G, T)$ that covers $r$. 
By Lemma~\ref{lem:connjoin2char}~\ref{item:nont} and \ref{item:tailnei}, 
$(G, T)$ is obtained from $(\initialgtr{G}{T}{r}, T)$ and $(G-\initialgtr{G}{T}{r}, \emptyset)$ 
by joining $\agtr{G}{T}{r}$ and $G-\initialgtr{G}{T}{r}$ with some edges. 

Let $\mathcal{K} := \{ K \in \laycompall{G}{T}{r} : V(K) \subseteq V(\initialgtr{G}{T}{r}) \}$,  
and  for each $K \in \mathcal{K}$, let $r_K := r$ if $K = \initialgtr{G}{T}{r}$,  and let $r_K$ be the $F$-root of $K$ otherwise. 
By Lemma~\ref{lem:connjoin2char} \ref{item:nonq}, 
we have $\mathcal{K} \subseteq \laycompall{G}{T}{r}\cap \qcompall{G}{T}{r}$.  
Let $\mathcal{K}' := \{ K \in \mathcal{K}: \dk{G}{T}{K} \neq \emptyset\} = \{ K \in \mathcal{K} : |V(K)| > 1\}$.  
Note $\initialgtr{G}{T}{r} \in \mathcal{K}'$ by Lemma~\ref{lem:connjoin2char} \ref{item:nontri}.

\begin{pclaim}  \label{claim:char:ind} 
The following properties hold for every $K\in \mathcal{K}'$. 
\begin{rmenum} 
\item \label{item:char:ind:gluesum} $(G, T)_F[K]$ is a gluing sum  $(G_K, T_K; S_K, F_K)\oplus \{ (G_L, T_L; A_L, r_L) : L \in \dkconn{G}{T}{K} \}$, 
where $(G_K, T_K) = (G, T)_F[K]/\dkconn{G}{T}{K}$,  
$S_K = \llbracket \dkconn{G}{T}{K} \rrbracket$, 
$F_K$ is the set of edges that correspond to the $F$-beams of all members of $\dkconn{G}{T}{K}$, 
$(G_L, T_L) = (G, T)_F[L]$, and $A_L = \ak{G}{T}{L}$. 
\item \label{item:char:ind:rake} $(G_K, T_K)$ is a member of $\mathcal{R}(r_K, \llbracket \dkconn{G}{T}{K} \rrbracket)$, and $F_K = \{ r_K[L] : L \in \dkconn{G}{T}{K} \}$. 
\end{rmenum} 
\end{pclaim} 
\begin{proof}[Proof of Claim~\ref{claim:char:ind}] 
Proposition~\ref{prop:neidist} implies $\parNei{G}{\ak{G}{T}{K}}\cap \dk{G}{T}{K} \subseteq \bigcup \{ \ak{G}{T}{L} : L \in \dkconn{G}{T}{K}\}$. 
Hence, \ref{item:char:ind:gluesum} holds. 
Proposition~\ref{prop:rakechar} and Lemma~\ref{lem:connjoin2char} \ref{item:rake}  imply $(G_K, T_K) \in \mathcal{R}(r_K, \llbracket \dkconn{G}{T}{K} \rrbracket)$. 
Lemma~\ref{lem:connjoin2char} \ref{item:joinproj} implies $F_K = \{ r_K[L] : L \in \dkconn{G}{T}{K} \}$. 
Thus, \ref{item:char:ind:rake} is proved.  
\end{proof}

We prove $(G, T)_F[K] \in \mathcal{P}(r_K, \ak{G}{T}{K})$ for every $K \in \mathcal{K}'$ by induction on $|\dk{G}{T}{K}|$. 
If $|V(K)| = 1$, then $(G, T)_F[K]$ is a trivial graft.  
Assume that $(G, T)_F[L] \in \mathcal{P}(r_L, \ak{G}{T}{L})$ for every $L \in \mathcal{K}'$ with $V(L) \subsetneq V(K)$.  
Then, by Claim~\ref{claim:char:ind}, we have $(G, T)_F[K] \in \mathcal{P}(r_K, \ak{G}{T}{K})$. 
Thus, we obtain $(\initialgtr{G}{T}{r}, T) \in \mathcal{P}(r, \agtr{G}{T}{r})$, and  the statement \ref{item:char:p} is proved.

Conversely, if \ref{item:char:p} holds, then Lemmas~\ref{lem:biprimal2const} and \ref{lem:tail} imply \ref{item:char:join}. 
Hence, the theorem follows. 
\end{proof}

Since $\mathcal{R}(r, B)$ is inductively constructed from scratch,  
so is $\mathcal{P}(r, A)$. 
Consequently, Theorem~\ref{thm:char} provides a construction of grafts with a connected minimum join from scratch.

\section{Algorithm for Recognizing Grafts with Connected Minimum Joins}  \label{sec:alg}

\subsection{Algorithmic Preliminaries}  \label{sec:alg:pre}

In this section, unless specified otherwise, $n$ and $m$ denote the number of vertices and edges of the input graph. 
Seb\H{o} and  Tannier~\cite{10.1007/3-540-45535-3_30, sebHo2004metric} first presented a polynomial-time algorithm 
for deciding whether a given graft has a connected minimum join. 
They derived this result as an application of their studies on metrics on graphs. 

\begin{theorem}[Seb\H{o} \&  Tannier~\cite{10.1007/3-540-45535-3_30, sebHo2004metric}]  \label{thm:sebot} 
For a graft $(G, T)$, 
whether $(G, T)$ has a connected minimum join can be decided in $O(n^3)$ time.   
If the graft has one, then it can also be computed in the same time bound. 
\end{theorem} 

We present a new algorithm that solves the same problem in $O(n(m + n\log n))$ time. 
In the remaining part of Section~\ref{sec:alg:pre}, we provide preliminary algorithmic results employed in our algorithm.

\begin{theorem}[Gabow~\cite{10.1145/3183369}]  \label{thm:gabow} 
Given a graft $(G, T)$, 
a minimum join of $(G, T)$ can be computed in $O(|T|(m + n\log n))$ time. 
\end{theorem}

\begin{definition} 
Let $G$ be a graph. 
A mapping $w: E(G) \rightarrow \mathbb{R}$ is called an edge weight of $G$.   
An edge weight $w$ is said to be {\em conservative} 
if $\sum_{e\in E(C)} w(e) \ge 0$ for every circuit $C$ of $G$.   
For a conservative edge weight $w$ of $G$, 
the {\em $w$-distance} between two vertices $x$ and $y$ 
is the minimum value of $\sum_{e\in E(P)} w(e)$ where $P$ is taken over all paths between $x$ and $y$. 
\end{definition}

Gabow and Sankowski~\cite{doi:10.1137/16M1106225} proposed an algorithm for constructing a data structure that efficiently stores
the solutions of the single-source all-sink shortest path problem in undirected graphs.  
Their work includes the following theorem as  part of their results.

\begin{theorem}[Gabow \& Sankowski~\cite{doi:10.1137/16M1106225}] \label{thm:gabows} 
Given a graph $G$, a conservative edge weight $w: E(G)\rightarrow \mathbb{Z}$,  and $r\in V(G)$,  
a mapping $\lambda_w: V(G) \rightarrow \mathbb{Z}$ 
that, given $x \in V(G)$,  returns the $w$-distance between $r$ and $x$ in $O(1)$ time 
can be computed in $O( \min\{ n(m + n\log n), m\sqrt{n}\log (nW) \})$ time,  where $W := \max \{ | w(e) |: e\in E(G) \}$.    
\end{theorem}

By letting $W = 1$ in Theorem~\ref{thm:gabows}, 
we obtain Proposition~\ref{prop:distalg}.

\begin{proposition}  \label{prop:distalg} 
Given a connected graft $(G, T)$, a minimum $T$-join $F$, and $r\in V(G)$, 
$\distgtf{G}{T}{F}{r}{x}$ for all $x\in V(G)$ can be computed in $O( m \sqrt{n}\log n)$ time. 
\end{proposition}

It is rather easily observed that the distance decomposition can be computed recursively from the minimum-distance side 
by using the connected component decomposition algorithm.

\begin{proposition}  \label{prop:seboalg} 
Given a connected graft $(G, T)$, $r \in V(G)$,  
and $\distgt{G}{T}{r}{x}$ for every $x\in V(G)$,  
the distance decomposition of $(G, T)$ with root  $r$ can be computed in $O(m)$ time. 
More precisely, 
for every $i\in \interval{G}{T}{r}$, 
the set $\levelgtr{G}{T}{r}{i}$, together with the partitions 
$\alpart{i}  := \{ \ak{G}{T}{K}: K \in \laycompgtr{G}{T}{r}{i} \}$ 
and $\aqpart{i} := \{ \ak{G}{T}{K}: K \in \qcompgtr{G}{T}{r}{i} \}$ of $\levelgtr{G}{T}{r}{i}$   
can be computed in $O(m)$ time so that the following properties also hold: 
\begin{rmenum} 
\item  \label{item:seboalg:f} 
Given $\ak{G}{T}{K} \in \alpart{i}$, where $K \in \laycompgtr{G}{T}{r}{i}$,   
the family $\{ \ak{G}{T}{L} : L \in \qkconn{G}{T}{K} \} \subseteq \aqpart{i}$ can be returned in $O(| \qkconn{G}{T}{K} |)$ time.   
\item \label{item:seboalg:g} 
Given $\ak{G}{T}{K} \in \aqpart{i}$, where $K \in \qcompgtr{G}{T}{r}{i}$, 
whether $\dk{G}{T}{K}$ is $\emptyset$ can be determined in $O(1)$ time. 
If $\dk{G}{T}{K} \neq \emptyset$, then 
 $\{ \ak{G}{T}{L} : L \in \dkconn{G}{T}{K} \}$ can be computed in $O( |\dkconn{G}{T}{K}| )$ time.  
\end{rmenum} 
\end{proposition}

\subsection{Algorithms for Connected Minimum Joins} 

\subsubsection{Eligible System and the Set of Heads}  \label{sec:alg:algconn:eligible} 
 
In this section, we introduce the concepts of an eligible system and the mapping $h$ 
and derive Lemma~\ref{lem:h2char} to be used in deriving our algorithm for a connected minimum join.

\begin{definition} 
Let $(G, T)$ be a connected graft, and let $r\in V(G)$. 
We say that $(G, T; r)$ is {\em eligible} if it satisfies the following properties: 
\begin{rmenum} 
\item $T\neq \emptyset$ and $T \subseteq V(\initialgtr{G}{T}{r})$. 
\item $| \qkconn{G}{T}{K} | = 1$ for every $K \in \noncapall{G}{T}{r} \cup \{ \initialgtr{G}{T}{r} \}$. 
\end{rmenum} 
\end{definition}

From Lemma~\ref{lem:connjoin2char}, if a graft $(G, T)$ has a connected minimum join covering $r$, 
then $(G, T; r)$ is eligible.

\begin{definition} 
Let $(G, T)$ be a connected graft, and let $r\in V(G)$ be such that $(G, T; r)$ is eligible. 
Define a mapping $h: \noncapall{G}{T}{r} \cup \{ \initialgtr{G}{T}{r} \} \rightarrow 2^{V(G)}$ as follows: 
\begin{rmenum} 
\item For $K \in \laycompall{G}{T}{r}$ with $\dk{G}{T}{K} = \emptyset$,  let $h(K) := V(K)$; and, 
\item for $K \in \noncapall{G}{T}{r}$ with  $\dk{G}{T}{K} \neq  \emptyset$ (resp. $K \in \laycompall{G}{T}{r} \setminus \noncapall{G}{T}{r}$),  
a vertex $v \in V(G)$ is in $h(K)$    if 
\begin{rmenum}
\item $v\in \ak{G}{T}{K}$ holds,  
\item $E_G[v, h(L)] \neq \emptyset$ for every $L \in \dkconn{G}{T}{K}$,   and 
\item  $|\{v\} \cap T | + |\dkconn{G}{T}{K}|$ is odd (resp. even).  
\end{rmenum} 
\end{rmenum} 
\end{definition}

Lemma~\ref{lem:rootalt} can be observed by Theorem~\ref{thm:sebo:path} and Proposition~\ref{prop:distalt}.

\begin{lemma}  \label{lem:rootalt} 
Let $(G, T)$ be a connected graft, let $F$ be a minimum join of $(G, T)$, and let $r\in V(G)$.  
Let $K \in \noncapall{G}{T}{r} \cup \{ \initialgtr{G}{T}{r} \}$. 
Let $r_K$ be $r$ if $K = \initialgtr{G}{T}{r}$, and the $F$-root of $K$ otherwise. 
Let $T_K := T\cap V(K)$. 
For every $v\in \ak{G}{T}{K}$,  
let $T_v$ be $T_K \Delta \{r \} \Delta \{ v \}$ if $K = \initialgtr{G}{T}{r}$, 
and $T_K \Delta \{v\}$ otherwise.  
Then, the following properties hold:  
\begin{rmenum} 
\item \label{item:rootalt:joinalt} 
The graft $(K, T_v)$ satisfies $\nu(K, T_v) = |F[K]|$.  
\item \label{item:rootalt:distalt}  
$\distgt{K}{T_v}{v}{x} = \distp{ (G, T)_F[K] }{r_K}{x}$ for every $x \in V(K)$.  
Accordingly, $\ak{G}{T}{K} = \{ x \in V(K) :  \distgt{K}{T_v }{v}{x} = 0 \}$, 
and $\dk{G}{T}{K} = \{ x \in V(K) :  \distgt{K}{T_v }{v}{x} < 0 \}$.  
\end{rmenum} 
\end{lemma} 
\begin{proof}By Theorem~\ref{thm:sebo:path},  
$\ak{G}{T}{K}$ (resp. $\dk{G}{T}{K}$)  
is the set of vertices in  $(G, T)_F[K]$ 
whose $F[K]$-distance from $r_K$ is $0$ (resp. negative).  
Thus, the statements hold if $v = r_K$.  Assume $v\neq r_K$ in the following. 
By Proposition~\ref{prop:distalt}, Theorem~\ref{thm:sebo:path} further implies  that 
$\nu(K, T_v) = \nu( (G, T)_F[K] )  = |F[K]|$,   
and $\distgt{K}{T_v}{v}{x} = \distp{(G, T)_F[K]}{r_K}{x}$ for every $x \in V(K)$. 
Thus, Statements \ref{item:rootalt:joinalt} and \ref{item:rootalt:distalt} follow. 
\end{proof}

Lemma~\ref{lem:h2char} follows from Lemma~\ref{lem:rootalt}, together with Theorem~\ref{thm:sebo:beam}.

\begin{lemma}  \label{lem:h2char} 
Let $(G, T)$ be a connected graft and let $r\in V(G)$ be such that $(G, T; r)$ is eligible. Let $F$ be a minimum join of $(G, T)$ that covers $r$.  
For each $K \in \laycompall{G}{T}{r}$, let $T_K$ denote $T\cap V(K)$. 
Then, the following properties hold: 
\begin{rmenum} 
\item \label{item:h2char:noncap} For $K \in \noncapall{G}{T}{r}$ with  $\dk{G}{T}{K} \neq  \emptyset$,  
a vertex $v \in \ak{G}{T}{K}$ is in $h(K)$  if and only if it can be covered by a connected minimum join of $(K, T_K \Delta \{v\})$. 
\item \label{item:h2char:initial}  
For $K = \initialgtr{G}{T}{r}$, 
a vertex $v \in \ak{G}{T}{K}$ is in $h(K)$  if and only if it can be covered by a connected minimum join of $(G, T)$.  
\end{rmenum} 
\end{lemma} 
\begin{proof}[Proof of Lemma~\ref{lem:h2char}] Let $K \in \noncapall{G}{T}{r} \cup \{\initialgtr{G}{T}{r} \}$.  
We prove the lemma by induction on $|\dk{G}{T}{K} |$.  
If $\dk{G}{T}{K}  = \emptyset$, then the statements are trivially true.   
Assume that $|\dk{G}{T}{K} | > 0$ and that the statements hold for every $K$ with smaller $|\dk{G}{T}{K} |$. 
First, consider the case  $K \in \noncapall{G}{T}{r}$ with  $\dk{G}{T}{K} \neq  \emptyset$. 

\begin{pclaim} \label{claim:h2char:h2cover} 
Every vertex $v$ in $h(K)$ can be covered by a connected minimum join of $(K, T_K \Delta \{v\})$. 
\end{pclaim} 
\begin{proof}[Proof of Claim~\ref{claim:h2char:h2cover}] 
Let $v\in h(K)$.  
For each $L \in \dkconn{G}{T}{K}$,  let $v_L \in h(L)$ be a vertex with $vv_L \in E(G)$; such $v_L$ exists from the definition of $h(K)$. 
By the induction hypothesis,  
if $\dk{G}{T}{L}\neq \emptyset$, then there is a connected minimum join $F_L$ of $(L,  T_L \Delta \{v_L\})$ that covers $v_L$; 
if $\dk{G}{T}{L} = \emptyset$, then let $F_L := \emptyset$.  
Let $F^v := \{ vv_L : L \in \dkconn{G}{T}{K} \} \cup \bigcup_{L \in \dkconn{G}{T}{K} } F_L$.

It is clear that $F^v$ is a connected join of $(K, T_K \Delta \{v\})$. 
To prove its minimality,  we only need to prove $|F^v| = |F[K]|$, owing to Lemma~\ref{lem:rootalt} \ref{item:rootalt:joinalt}. 
By Theorem~\ref{thm:sebo:beam}~\ref{item:sebo:beam:noncap}, 
we have $|F[K]| = |\dkconn{G}{T}{K}| + |\bigcup_{L \in \dkconn{G}{T}{K}} F[L]|$. 
It is clear that $|\dkconn{G}{T}{K}| = |\{ vv_L : L \in \dkconn{G}{T}{K} \}|$, whereas 
$|F_L| = | F[L] |$ holds for every $L \in \dkconn{G}{T}{K}$, according to Lemma~\ref{lem:rootalt} \ref{item:rootalt:joinalt}. 
Hence, we obtain $|F^v| = |F[K]|$. 
Consequently, $F^v$ is a connected minimum join of $(K, T_K \Delta \{v\})$. 
The claim is proved.  
\end{proof}

\begin{pclaim} \label{claim:h2char:cover2h} 
If a vertex $v\in \ak{G}{T}{K}$ is covered by a connected minimum join of $(K, T_K \Delta \{v\})$, 
then $v$ is an element of $h(K)$. 
\end{pclaim} 
\begin{proof}[Proof of Claim~\ref{claim:h2char:cover2h}] 
Let $v\in \ak{G}{T}{K}$, and assume that $v$ is covered by a connected minimum join $F^v$ of $(K, T_K \Delta \{v\})$.  
By Lemma~\ref{lem:rootalt}~\ref{item:rootalt:distalt}, 
$\ak{G}{T}{K} = \{ x\in V(K): \distgt{K}{T_K\Delta \{v\}}{v}{x} \allowbreak = 0\}$ and 
$\dk{G}{T}{K} = \{ x\in V(K): \distgt{K}{T_K\Delta \{v\}}{v}{x} < 0 \}$; 
note that $(K, T_K \Delta \{v\})$ is a primal graft with respect to $v$, and $K = \initialgtr{K}{T_K\Delta \{v\}}{v}$. 
Thus, by Lemma~\ref{lem:connjoin2char}~\ref{item:joinproj},  
$F^v$ is the union of two sets $\{ vv_L: L \in \dkconn{G}{T}{K} \}$ and $\bigcup \{ F_L:  L \in \dkconn{G}{T}{K} \mbox{ with } \dk{G}{T}{L}\neq \emptyset \}$;  here,  $v_L$ is a vertex in $\ak{G}{T}{L}$, and $F_L$ is a connected minimum join of $(L,  T_L \Delta \{v_L\})$  that covers $v_L$.  
The induction hypothesis implies that $v_L \in h(L)$ for every $L \in \dkconn{G}{T}{K}$. 
Thus, $v\in h(K)$ follows. 
\end{proof}

Claims~\ref{claim:h2char:h2cover} and \ref{claim:h2char:cover2h} imply \ref{item:h2char:noncap}. 
It can be proved in a similar way that 
$h(\initialgtr{G}{T}{r})$ is the set of vertices that can be covered by a connected minimum join of $(G, T)[\initialgtr{G}{T}{r}]$.  
Further, $T\subseteq V(\initialgtr{G}{T}{r})$ and Theorem~\ref{thm:sebo:beam}~\ref{item:sebo:beam:cap} imply 
 that every minimum join of $(G, T)[\initialgtr{G}{T}{r}]$ is a minimum join of $(G, T)$, and vice versa. 
Thus, the statement \ref{item:h2char:initial} follows.  
This completes the proof. 
\end{proof}

\subsubsection{Algorithmic Results} 

In this section, we present  a new algorithm in Theorem~\ref{thm:alg} for deciding whether a given graft has a connected minimum join.   
We first establish Lemmas~\ref{lem:eligible} and \ref{lem:h} and then use these lemmas to prove Theorem~\ref{thm:alg}.

\begin{lemma} \label{lem:eligible} 
Given a connected graft $(G, T)$, a minimum join $F$, a vertex $r\in V(G)$, 
the values $\distgtf{G}{T}{F}{r}{x}$ for all $x\in V(G)$, and the distance decomposition of $(G, T)$ with root $r$, 
whether $(G, T; r)$ is eligible can be determined in $O(m)$ time. 
\end{lemma} 
\begin{proof}[Proof of Lemma~\ref{lem:eligible}]  If $T = \emptyset$, then the statement obviously holds. Hence, assume $T\neq \emptyset$ in observing the following three claims: 

\begin{pclaim} \label{claim:eligible:noncap2initial} 
If $T\subseteq V(\initialgtr{G}{T}{r})$ holds,  
then every member of $\noncapall{G}{T}{r}$ is a subgraph of $\initialgtr{G}{T}{r}$. 
\end{pclaim} 
\begin{proof}[Proof of Claim~\ref{claim:eligible:noncap2initial}]  
Assume that there exists $K \in \noncapall{G}{T}{r}$ that is not a subgraph of $\initialgtr{G}{T}{r}$. 
Theorem~\ref{thm:sebo:beam} implies $\parcut{G}{\initialgtr{G}{T}{r}} \cap F = \emptyset$ 
and $\parcut{G}{K} \cap F  \neq \emptyset$. 
Hence, the subgraph of $G$ induced by $F$ has a connected component $C$ that is a subgraph of $G - \initialgtr{G}{T}{r}$. 
By Lemma~\ref{lem:circuit}, $C$ is a tree. 
Hence, there exists $v\in V(C)$ with $|\parcut{G}{v} \cap F| = 1$, which implies $v\in T \setminus V(\initialgtr{G}{T}{r})$. 
Therefore, $T \not\subseteq V(\initialgtr{G}{T}{r})$ follows.  
The claim is proved. 
\end{proof}

From Claim~\ref{claim:eligible:noncap2initial}, we prove Claim~\ref{claim:eligible:initial}.

\begin{pclaim}  \label{claim:eligible:initial} 
$T \subseteq V(\initialgtr{G}{T}{r})$ holds 
if and only if  $\laycompgtr{G}{T}{r}{0} = \{ \initialgtr{G}{T}{r} \}$ and $T \subseteq \laylegtr{G}{T}{r}{0}$ hold. 
\end{pclaim} 
\begin{proof}[Proof of Claim~\ref{claim:eligible:initial}] 
First, we prove the sufficiency part.  
Assume $T \subseteq V(\initialgtr{G}{T}{r})$; then by Claim~\ref{claim:eligible:noncap2initial}, 
we have $\laycompgtr{G}{T}{r}{0} = \{ \initialgtr{G}{T}{r} \}$. 
The remaining claim $T \subseteq \laylegtr{G}{T}{r}{0}$ is obvious from $V(\initialgtr{G}{T}{r}) \subseteq \laylegtr{G}{T}{r}{0}$.  
Thus, the sufficiency part is proved. 

We next prove the necessity part. 
Assume $\laycompgtr{G}{T}{r}{0} = \{ \initialgtr{G}{T}{r} \}$ and $T \subseteq \laylegtr{G}{T}{r}{0}$; 
the first condition is equivalent to $\laylegtr{G}{T}{r}{0} = V(\initialgtr{G}{T}{r})$. 
Hence, the second condition implies $T\subseteq V(\initialgtr{G}{T}{r})$.  
The necessity part is proved. 
\end{proof} 

\begin{pclaim} \label{claim:eligible:negcomp2initial} 
Assume $T \subseteq V(\initialgtr{G}{T}{r})$. 
Then, $\noncapall{G}{T}{r} \cup \{ \initialgtr{G}{T}{r} \} = \bigcup \{ \laycompgtr{G}{T}{r}{i} : i \in \interval{G}{T}{r}\cap \mathbb{Z}_{\le 0 } \}$.  
\end{pclaim} 
\begin{proof}[Proof of Claim~\ref{claim:eligible:negcomp2initial}]  
Note $\bigcup \{ \laycompgtr{G}{T}{r}{i} : i \in \interval{G}{T}{r}\cap \mathbb{Z}_{\le 0 } \}  
= \bigcup \{ \laycompgtr{G}{T}{r}{i} : i \in \interval{G}{T}{r}\cap \mathbb{Z}_{< 0 } \}  \cup  \laycompgtr{G}{T}{r}{0}$.  
By Claim~\ref{claim:eligible:initial},  
we have $\laycompgtr{G}{T}{r}{0} = \{ \initialgtr{G}{T}{r} \}$. 
Hence, we only need to prove  
$\bigcup \{ \laycompgtr{G}{T}{r}{i} : i \in \interval{G}{T}{r}\cap \mathbb{Z}_{< 0 } \}  = \noncapall{G}{T}{r}$. 
It is clear that any $\laycompgtr{G}{T}{r}{i}$ with $i \in \interval{G}{T}{r}\cap \mathbb{Z}_{< 0 }$  is a subfamily of $\noncapall{G}{T}{r}$. 
Additionally, by Claim~\ref{claim:eligible:noncap2initial}, 
if $K$ is a member of $\noncapall{G}{T}{r}$, then $V(K)\subseteq V(\initialgtr{G}{T}{r})$ holds, 
and $K\in \laycompgtr{G}{T}{r}{i}$ follows for some $i\in \interval{G}{T}{r}$ with $i <0$. 
Thus, the claim is proved. 
\end{proof} 
By Claim~\ref{claim:eligible:initial}, 
for examining whether $(G, T; r)$ with $T\neq \emptyset$ satisfies the condition  $T \subseteq V(\initialgtr{G}{T}{r})$, 
we only need 
to examine whether $|\laycompgtr{G}{T}{r}{0}| = 1$ and $\distgtf{G}{T}{F}{r}{x} \le 0$ for every $x\in T$.  
Owing to Claim~\ref{claim:eligible:negcomp2initial}, 
once $T\subseteq V(\initialgtr{G}{T}{r})$ is confirmed, 
whether $|\qkconn{G}{T}{K}| =1 $ holds for every $K \in \noncapall{G}{T}{r} \cup \{ \initialgtr{G}{T}{r}\}$ 
can be confirmed 
by examining whether $|\qkconn{G}{T}{K}| =1 $ holds for every $K \in \bigcup \{ \laycompgtr{G}{T}{r}{i} : i \in \interval{G}{T}{r}\cap \mathbb{Z}_{\le 0 } \}$. 
As such, it easily follows 
that the following procedure returns  Yes if $(G, T; r)$ is eligible and No otherwise.  

\begin{algorithmic}[1]
\If{ $T = \emptyset$ } \label{if:tempty}  \Return No; \EndIf  \label{endif:tempty} 
\If{ $|\laycompgtr{G}{T}{r}{0}| \neq 1$}  \label{if:initial} \Return No; \EndIf  \label{endif:initial} 
\ForAll{ $v\in T$ }  \label{for:t2initial} 
\If{ $\distgtf{G}{T}{F}{r}{v} > 0$ } \Return No; \EndIf  
\EndFor  \label{endfor:t2initial} 
\ForAll{ $\ak{G}{T}{K}$ where $K \in \bigcup \{ \laycompgtr{G}{T}{r}{i} : i \in \interval{G}{T}{r}\cap \mathbb{Z}_{\le 0 } \}$ }   \label{for:q2uni} 
\If{ $| \{ \ak{G}{T}{L} : L \in \qkconn{G}{T}{K} \}| \neq 1$} \Return No; \EndIf 
\EndFor \label{endfor:q2uni} 
\State \Return Yes. 
\end{algorithmic} 

Clearly, Lines~\ref{if:tempty} to \ref{endif:tempty} take constant time.  
Lines~\ref{if:initial} to \ref{endif:initial} take linear time. 
Lines~\ref{for:t2initial} to \ref{endfor:t2initial}  take linear time. 
By Proposition~\ref{prop:seboalg}, Lines~\ref{for:q2uni} to \ref{endfor:q2uni}  
take linear time. 
Thus, the entire procedure runs in linear time. 
This proves the lemma. 
\end{proof}

Lemma~\ref{lem:eligible2comp} follows directly from the proof of Lemma~\ref{lem:eligible}.

\begin{lemma} \label{lem:eligible2comp} 
Let $(G, T)$ be a connected graft, and let $r\in V(G)$. 
If $(G, T; r)$ is eligible, 
then $\noncapall{G}{T}{r} \cup \{ \initialgtr{G}{T}{r} \} = \bigcup \{ \laycompgtr{G}{T}{r}{i} : i\in\interval{G}{T}{r} \cap \mathbb{Z}_{\le 0} \}$. 
\end{lemma}

\begin{lemma}  \label{lem:h} 
Let $(G, T)$ be a connected graft and $r\in V(G)$ be such that $(G, T; r)$ is eligible. 
Given the distance decomposition of $(G, T)$ with respect to $r$, 
the set $h(\initialgtr{G}{T}{r})$ can be computed in $O(m)$ time. 
\end{lemma} 
\begin{proof}[Proof of Lemma~\ref{lem:h}]
By the definition of $h$, 
we can compute $h(\initialgtr{G}{T}{r})$ 
recursively from the minimum-distance side along the distance decomposition. 
More precisely, $h(\initialgtr{G}{T}{r})$ can be obtained by the following procedure. 
\begin{algorithmic}[1]
\State Initialize $h$ as $h(K) := \emptyset$ for every $K \in \noncapall{G}{T}{r} \cup \{\initialgtr{G}{T}{r}\}$;  \label{line:init_h} 
\ForAll{$i = \min \interval{G}{T}{r}$ to $i = 0$}  \label{line:forall_i} 
\ForAll{$K \in \laycompgtr{G}{T}{r}{i}$} \label{line:forall_k} 
\If{ $\dk{G}{T}{K}  = \emptyset$ } \label{line:if_edge} 
\State $h(K) := V(K)$ \label{line:edge}; {\bf continue}; \Comment{The base case.} 
\EndIf   \label{line:endif_edge} 
\State initialize $owner: V(G)\to \mathcal{K}^*$ as an empty function; 
\State let $h'_K := \ak{G}{T}{K} \cap \bigcap_{L \in \dkconn{G}{T}{K}} \parNei{G}{h(L)}$;  \label{line:neicap} 
\State let $d \in \{0, 1\}$ be such that $d \equiv |\dkconn{G}{T}{K}| \pmod 2$; \label{line:d} 
\ForAll{$v \in h'_K$} \label{line:forall_H} \Comment{Checking the parity condition for $v$.} 
\State let $t := |\{ v \} \cap T|$;  \label{line:t} 
\If{$i < 0$ $\wedge$ $t + d$ is odd} \State $h(K) := h(K) \cup \{v\}$; \ElsIf{$i = 0$ $\wedge$ $t + d$ is even}  \State $h(K) := h(K) \cup \{v\}$; \EndIf
\EndFor \label{line:endfor_H} 
\EndFor
\EndFor 
\State \Return $h$
\end{algorithmic} 
Under Lemma~\ref{lem:eligible2comp}, 
let $\mathcal{K} :=  \noncapall{G}{T}{r}  \cup \{\initialgtr{G}{T}{r} \} = \bigcup \{ \laycompgtr{G}{T}{r}{i}: i\in \interval{G}{T}{r} \cap \mathbb{Z}_{\le 0} \}$. 
Note that the loops of Lines~\ref{line:forall_i} and \ref{line:forall_k} 
together examine every $K \in \mathcal{K}$. 
We first prove the correctness of the procedure 
as in Claim~\ref{claim:h:h}.

\begin{pclaim}  \label{claim:h:h}
The procedure outputs $h(K)$ for every $K \in \mathcal{K}$. 
\end{pclaim} 
\begin{proof}[Proof of Claim~\ref{claim:h:h}] 
The claim can be confirmed by the induction on $|\dk{G}{T}{K}|$. 
If $|\dk{G}{T}{K}| = 0$, then Line~\ref{line:edge} obviously returns $h(K)$. 
Assume $|\dk{G}{T}{K}| > 0$, and assume that the claim holds for every case where $|\dk{G}{T}{K}|$ is less; 
consequently, the claim holds for every $L \in \dkconn{G}{T}{K}$.   
Accordingly, by Line~\ref{line:neicap}, $h_K'$ is the set of vertices that have a neighbor in $h(L)$ for every $L \in \dkconn{G}{T}{K}$. 
Lines~\ref{line:forall_H} to \ref{line:endfor_H} then check the remaining condition for the elements in $h'_K$ to be a member of $h(K)$ 
and obtain $h(K)$. 
Thus, the correctness of the procedure follows. 
\end{proof}

We next prove the time complexity. 
Note $|\mathcal{K}| = O(n)$ in the following. 
Line~\ref{line:init_h} takes $O(n)$ time.  
The for-loops of Lines~\ref{line:forall_i} and \ref{line:forall_k} together repeat $|\mathcal{K}|$ times.  
Also, for every $i\in \interval{G}{T}{r} \cap \mathbb{Z}_{\le 0}$, 
$\laycompgtr{G}{T}{r}{i}$ can be obtained in $O(|\laycompgtr{G}{T}{r}{i}|)$ time, owing to Proposition~\ref{prop:seboalg}. 
Hence, choosing every $K \in \mathcal{K}$ for the repetition takes $O(n)$ time in total.

Lines~\ref{line:if_edge} to \ref{line:endif_edge} cost $O(1)$ per execution; hence they cost $O(n)$ time in total during the whole procedure. 
The operations from Line~\ref{line:neicap} onward are performed for every $K \in \mathcal{K}^*$, 
where $\mathcal{K}^* := \{ K \in \mathcal{K} : \dk{G}{T}{K} \neq \emptyset \}$.

\begin{pclaim} \label{claim:h:neicap}
Line~\ref{line:neicap} can be performed in $O(n+m)$ time in total during the whole procedure. 
\end{pclaim} 
\begin{proof}[Proof of Claim~\ref{claim:h:neicap}] 
We prove that  $h'_K$ can be computed by the following procedure  for every $K \in \mathcal{K}^*$: 
\begin{algorithmic}[1] 
\State let $owner(u) := K$ for every $u \in \ak{G}{T}{K}$;  \label{line:ak} 
\State Initialize $cnt_K: \ak{G}{T}{K} \to \mathbb{Z}_{\ge 0}$ as $cnt_K := 0$; \label{line:cnt} 
\ForAll{ $L \in \dkconn{G}{T}{K}$ }  \label{forall:l} 
\State initialize $mk_L: \ak{G}{T}{K} \to \{ L \}$ as an empty function;  \label{line:mk} 
\ForAll{ $v\in h(L)$}  \label{forall:v} 
\ForAll{ $vw  \in E(G)$ }  \label{forall:w} 
\If{ $owner(w) = K   \wedge mk_L(w) = \perp$ }   \label{if:w} 
\State let $cnt_K(w) := cnt_K(w) + 1$   and $mk_L(w) := L$;  \label{line:add} 
\EndIf \label{endif:w} 
\EndFor \label{endfor:w} 
\EndFor \label{endfor:v} 
\EndFor \label{endfor:l} 
\State output $h_K'$ as $\{ w \in \ak{G}{T}{K} : cnt_K(w) = d\}$, where $d = |\dkconn{G}{T}{K}|$.  \label{line:output} 
\end{algorithmic}

The correctness of the procedure is clear. 
We prove the time complexity in the following. 
At Lines~\ref{line:ak} and \ref{line:add}, we treat the assigned value $K$ or $L$ as fresh tokens associated with $K \in \mathcal{K}^*$ or $L \in \dkconn{G}{T}{K}$. 
Line~\ref{line:ak} and Line~\ref{line:cnt} take $O(|\ak{G}{T}{K}|)$ time for each $K \in \mathcal{K}^*$ 
and thus take $O(n)$ time in total during the whole procedure. 
By Proposition~\ref{prop:seboalg},  $\dkconn{G}{T}{K}$ can be obtained in $O(|\dkconn{G}{T}{K}|)$ time if $\ak{G}{T}{K}$ is given; 
hence, Line~\ref{forall:l} takes $O(|\dkconn{G}{T}{K}|)$ time for each $K \in \mathcal{K}^*$. 
Line~\ref{line:mk} takes $O(1)$ time per execution and thus takes $O(|\dkconn{G}{T}{K}|)$ time for each $K \in \mathcal{K}^*$. 
Thus, Lines~\ref{forall:l} and \ref{line:mk} take $O(n)$ time in total during the whole procedure.

Line~\ref{forall:v} takes $O(|\ak{G}{T}{L}|)$ time for every $L \in \dkconn{G}{T}{K}$ 
and hence takes $O(n)$ time in total during the whole procedure because every vertex in $V(G)$ is chosen at most once by Line~\ref{forall:v}. 
During the whole procedure, every edge in $\initialgtr{G}{T}{r}$ is chosen at most twice by Line~\ref{forall:w}.  
Hence, Line~\ref{forall:w} takes $O(m)$ time in total during the whole procedure. 
Lines~\ref{if:w} to \ref{endif:w} take $O(1)$ time per execution and thus take $O(m)$ time in total during the whole procedure. 
Line~\ref{line:output} can be done in $O(|\ak{G}{T}{K}|)$ time 
and thus takes $O(n)$ time in total during the whole procedure. 
\end{proof}

By Proposition~\ref{prop:seboalg}, Line~\ref{line:d} costs $O(|\dkconn{G}{T}{K}|)$ time for every $K \in \mathcal{K}^*$. 
Consequently it costs $O(n)$ time during the whole procedure, 
because every member of $\mathcal{K}$ is accessed once by this operation.  
The for-loop of 
Line~\ref{line:forall_H}  repeats at most $|\ak{G}{T}{K}|$ times. 
Every line in this for-loop costs $O(1)$ time. 
Hence, Lines~\ref{line:forall_H} to \ref{line:endfor_H} cost 
$O(\sum_{K \in \mathcal{K}^* } |\ak{G}{T}{K}|)$ time during the whole procedure, 
which is $O(n)$ time  
because every vertex contributes by at most one to this summation.

Thus, the whole procedure runs in $O(n+m)$ time, 
which is $O(m)$ time because $G$ is connected. 
This completes the proof of the lemma. 
\end{proof}

\begin{algorithm}         \caption{Algorithm for deciding whether a graft has a connected minimum join} \label{alg:connjoin}
\begin{algorithmic}[1]        \Require a graft $(G, T)$;   
\Ensure Yes if $(G, T)$ has a connected minimum join, otherwise No.  
\If{ $T = \emptyset$ } \label{if:t} \Return No; \EndIf 
\State compute the connected components of $G$; 
\If{ $T$ is contained in a single connected component $C$ of $G$} 
replace $(G, T)$ by $(C, T)$; 
\Else \Return No; 
\EndIf  \label{endif:conn} 
\State \label{state:minjoin} compute a minimum join $F$ of $(G, T)$; 
\State \label{state:sebo} choose a vertex $r \in T$; compute the distance decomposition with  root $r$;  
\If{ $(G, T ; r)$ is not eligible} \label{if:eligible} \Return No; \EndIf \label{endif:eligible}  
\If{ $h(\initialgtr{G}{T}{r}) = \emptyset$ } \label{if:h} \State \Return No; 
\Else \State \Return Yes.  \EndIf \label{endif:h} 
\end{algorithmic}
\end{algorithm}

\begin{theorem}  \label{thm:alg}
Given a graft $(G, T)$, 
whether $(G, T)$ has a connected minimum join can be determined in $O(|T|(m + n\log n) + m\sqrt{n}\log n)$ time, 
which is bounded as $O(n(m + n\log n))$ time. 
A connected minimum join can also be computed in the same time complexity if any exists. 
\end{theorem} 
\begin{proof} By Lemmas~\ref{lem:connjoin2char} and \ref{lem:h2char}~\ref{item:h2char:initial}, the correctness of Algorithm~\ref{alg:connjoin} follows immediately. 
From Theorem~\ref{thm:gabow} and Propositions~\ref{prop:distalg} and  \ref{prop:seboalg},  
Lines~\ref{state:minjoin}  and   \ref{state:sebo}  take $O( |T|(m + n\log n) )$ and $O(m\sqrt{n}\log n)$ time, respectively. 
Lines~\ref{if:t} to \ref{endif:conn}, 
Lines~\ref{if:eligible} to \ref{endif:eligible},  and Lines~\ref{if:h} to \ref{endif:h} each take $O(n + m)$ time.  
Hence, Algorithm~\ref{alg:connjoin} runs in $O( |T|(m + n\log n) + m\sqrt{n}\log n )$ time.   

If $(G, T)$ has a connected minimum join, 
then Algorithm~\ref{alg:connjoin} leaves $h(\initialgtr{G}{T}{r}) \neq \emptyset$. 
It can easily be observed that,  by arbitrarily choosing a vertex $v$ from $h(\initialgtr{G}{T}{r})$, 
a connected minimum join covering $v$ can be constructed recursively in linear time. 
\end{proof}

If $|T|$ is significantly small in the input graft, then the time complexity of the algorithm can be bounded more tightly. 
Furthermore, 
when $|T|$ is significantly small, 
the minimum join problem can also be solved efficiently in $O(|T|^3)$ time 
via the classical reduction method~\cite{schrijver2003}  that uses the minimum weight perfect matching algorithm. 
In this case, Algorithm~\ref{alg:connjoin} can also be implemented to run in $O(|T|^3 + m\sqrt{n}\log n)$ time.

As observed in the proof of Theorem~\ref{thm:alg}, Algorithm~\ref{alg:connjoin} has two bottlenecks. 
One is the time required to compute an arbitrary minimum join $F$ of the graft, 
and the other is the time required to compute the $F$-distances between the root and all vertices. 
If the best known time complexities for these problems are improved,  the time complexity of our algorithm will improve accordingly.

\section*{Appendix: Proof of Proposition~\ref{prop:seboalg}}

\begin{proof}[Proof of Proposition~\ref{prop:seboalg}] 
For a graph $H$, we denote the set of connected components of $H$ by $\conn{H}$ in the following. 
We define $\alpha := \min \interval{G}{T}{r}$ and $\beta := \max \interval{G}{T}{r}$; 
that is, $\interval{G}{T}{r} = [\alpha, \beta]\cap \mathbb{Z}$ by Proposition~\ref{prop:neidist}. 
Define $\alpart{\alpha -1} := \emptyset$ for convenience. 
For each $i\in \interval{G}{T}{r}$, 
we define the graph $G_i$ as follows: 
\begin{itemize}
\item $V(G_i) = \alpart{i-1} \cup \levelgtr{G}{T}{r}{i}$, in which every $A \in \alpart{i-1}$ is a formal vertex, and 
\item $E(G_i) = \bigcup_{A \in \alpart{i-1}} E_G[A, \levelgtr{G}{T}{r}{i}] \cup E_G[\levelgtr{G}{T}{r}{i}]$; 
here every edge $xy\in E(G)$ between $x\in A$ and $y \in \levelgtr{G}{T}{r}{i}$ is identified as the edge $Ay \in E(G_i)$. 
\end{itemize}  
Also, we define $G_i' := G_i - E_G[\levelgtr{G}{T}{r}{i}]$.  
We define $f$ as a mapping defined on $\bigcup_{i=\alpha}^{\beta} \alpart{i}$ 
that returns  $\{ \ak{G}{T}{L} : L \in \qkconn{G}{T}{K} \}$ for every $\ak{G}{T}{K}$, where $K \in \laycompall{G}{T}{r}$. 
We also define $g$ as a mapping defined on $\bigcup_{i=\alpha}^{\beta} \aqpart{i}$ 
that returns  $\{ \ak{G}{T}{L} : L \in \dkconn{G}{T}{K}\}$ 
 for every $\ak{G}{T}{K}$, where $K \in \qcompall{G}{T}{r}$.

From the definition of the distance decomposition and Proposition~\ref{prop:neidist}, 
it can easily be observed that,  for every $i\in \interval{G}{T}{r}$, 
there is a naturally induced correspondence between $\laycompgtr{G}{T}{r}{i}$ and $\conn{G_i}$. 
Similarly, 
there is a natural correspondence between $\qcompgtr{G}{T}{r}{i}$ and $\conn{G_i'}$.  
Also, 
for every corresponding  pair of $K \in \laycompgtr{G}{T}{r}{i}$ and $C_K \in \conn{G_i}$, 
there is a natural correspondence between $\qkconn{G}{T}{K}$ and $\conn{C_K - E_G[\levelgtr{G}{T}{r}{i}]}$.

Therefore, 
we can inductively compute the desired partitions $\alpart{i}$ and $\aqpart{i}$, 
starting from the smallest $i\in \interval{G}{T}{r}$ and incrementing it to the maximum,  
by computing $G_i$ and $G_i'$ and their connected components. 
The mappings $f$ and $g$ can be constructed along the way. 
We propose the following procedure:

\begin{algorithmic}[1]
\State Initialize $f$ and $g$ as empty functions;  \label{line:fg} 
\State compute $\interval{G}{T}{r}$ and the family $\{ \levelgtr{G}{T}{r}{i} : i  \in \interval{G}{T}{r} \}$; \label{line:dist} 
\State let $\alpha := \min \interval{G}{T}{r}$ and let $\beta := \max \interval{G}{T}{r}$;  \label{line:minmax} 
\State initialize $\alpart{i} := \emptyset$ and $\aqpart{i} := \emptyset$ for every $i \in \interval{G}{T}{r}$;  \label{line:alaq} 
\State let $\alpart{\alpha - 1} := \emptyset$;  \label{line:bound} 
\For{$i = \alpha$ to $\beta$}   \label{forall:i} 
\State compute $G_i$; \label{line:gi} 
\State compute $\conn{G_i}$;  \label{line:ci} 
\ForAll{ $C \in \conn{G_i}$ }  \label{forall:c} 
\State let $A := V(C) \cap \levelgtr{G}{T}{r}{i}$, let $\alpart{i} := \alpart{i} \cup \{A\}$, and initialize $f(A):=\emptyset$;  \label{line:al}
\State compute $\conn{C - E_G[\levelgtr{G}{T}{r}{i}]}$;  \label{line:cp} 
\ForAll{$C' \in \conn{C - E_G[\levelgtr{G}{T}{r}{i}]}$}  \label{forall:cp} 
\State let $A' := V(C') \cap \levelgtr{G}{T}{r}{i}$ and $D' := V(C') \setminus \levelgtr{G}{T}{r}{i}$; \label{line:ap} 
\State let $f(A) := f(A) \cup \{A' \}$, and let $\aqpart{i} := \aqpart{i} \cup \{ A'\}$; 
\State let $g(A') := D'$;  \label{line:ap_map} 
\EndFor
\EndFor 
\EndFor
\State output $\alpart{i}$ and $\aqpart{i}$ for every $i\in \interval{G}{T}{r}$ and $f$ and $g$; stop. 
\end{algorithmic} 
The correctness of the procedure is clear from the correspondences between $\laycompgtr{G}{T}{r}{i}$ and $\conn{G_i}$ and so on. 
The mappings $f$ and $g$ satisfy the properties required by \ref{item:seboalg:f} and \ref{item:seboalg:g}, respectively.

We prove that the proposed procedure runs in $O(m)$ time in the following. 
For each  $i\in \interval{G}{T}{r}$,  let $n_i := |V(G_i)|$ and let $m_i := |E(G_i)|$. 
Note $O(n+m) = O(m)$ as $G$ is connected. 
Note also $\sum_{i=\alpha}^{\beta} |\levelgtr{G}{T}{r}{i}| = n$.

\begin{pclaim}  \label{claim:linear} 
It holds that $\sum_{i=\alpha}^{\beta} n_i \le 2n$ and $\sum_{i=\alpha}^{\beta} m_i = m$.  
Consequently, $O(\sum_{i=\alpha}^{\beta} n_i+m_i) = O(n+ m) = O(m)$. 
\end{pclaim} 
\begin{proof} 
Define $\levelgtr{G}{T}{r}{\alpha -1} := \emptyset$ for convenience.  
Because $|\alpart{i-1}| \le |\levelgtr{G}{T}{r}{i-1}|$ for every $i\in \interval{G}{T}{r}$,   
we have $n_i \le  |\levelgtr{G}{T}{r}{i-1}| + |\levelgtr{G}{T}{r}{i}|$.  
Hence, 
$\sum_{i=\alpha}^{\beta} n_i \le \sum_{i=\alpha}^{\beta}  |\levelgtr{G}{T}{r}{i-1}| + |\levelgtr{G}{T}{r}{i}| 
\le 2 \sum_{i=\alpha}^{\beta}  |\levelgtr{G}{T}{r}{i}| = 2n$.

For every $i\in \interval{G}{T}{r}$, 
we also have 
$m_i = | E_G[\levelgtr{G}{T}{r}{i}, \levelgtr{G}{T}{r}{i-1}]| + |E_G[\levelgtr{G}{T}{r}{i}]|$, 
because 
$\bigcup_{A \in \alpart{i-1}} E_G[A, \levelgtr{G}{T}{r}{i}] = E_G[\levelgtr{G}{T}{r}{i}, \levelgtr{G}{T}{r}{i-1}]$.   
Hence, 
we have 
$\sum_{i=\alpha}^{\beta} m_i = \sum_{i=\alpha}^{\beta} | E_G[\levelgtr{G}{T}{r}{i}, \levelgtr{G}{T}{r}{i-1}]| + |E_G[\levelgtr{G}{T}{r}{i}]| = m$, 
because every edge of $G$ is counted exactly once in the summation of the middle expression by Proposition~\ref{prop:neidist}. 
This completes the proof of the claim. 
\end{proof}

Line~\ref{line:fg} takes $O(1)$ time. 
Lines~\ref{line:dist} and \ref{line:minmax} can be performed in $O(n)$ time. Line~\ref{line:bound} takes $O(1)$ time. 
Line~\ref{line:alaq} takes $O(n)$ time. 
Note $\interval{G}{T}{r} \subseteq [-n, n] \cap \mathbb{Z}$. 
Hence, we can compute the family $\{ \levelgtr{G}{T}{r}{i} : i  \in \interval{G}{T}{r} \}$  in $O(n)$ time 
by examining and allocating  each vertex in $V(G)$ one by one.

Line~\ref{line:gi} costs $O(n_i + m_i)$ time, 
so does Line~\ref{line:ci},  since the connected component decomposition of a graph can be computed in linear time. 
Hence, by Claim~\ref{claim:linear}, Lines~\ref{line:gi} and \ref{line:ci} take $O(m)$ time in total during the whole procedure.

For each fixed $i\in \interval{G}{T}{r}$, 
Line~\ref{line:al} costs $O(n_i)$ time in total over all iterations of the loop in Line~\ref{forall:c}.  
Consequently, Line~\ref{line:al} costs $O(n)$ time in total during the whole procedure.

For each fixed $i$, Line~\ref{line:cp} takes $O(n_i+m_i)$ time in total
over all iterations of the loop in Line~\ref{forall:c}, since it is equivalent to
computing $\conn{G_i'}$. 
Here, $|V(G_i')| = n_i$ and $| E(G_i')| \le m_i$. 
Therefore, 
by Claim~\ref{claim:linear}, Line~\ref{line:cp} takes $O(m)$ time in total during the whole procedure.

For each fixed $i$, the nested loops in Lines~\ref{forall:c} and \ref{forall:cp} iterate over all components of $G_i'$. 
For each fixed $i\in \interval{G}{T}{r}$,  
Lines~\ref{line:ap} to \ref{line:ap_map} take 
$O(|\levelgtr{G}{T}{r}{i}|+|\levelgtr{G}{T}{r}{i-1}|)$ time in total 
over all iterations of the loops in Lines~\ref{forall:c} and \ref{forall:cp}, 
because each vertex in 
$\levelgtr{G}{T}{r}{i} \cup \levelgtr{G}{T}{r}{i-1}$ 
is accessed only a constant number of times. 
Therefore, 
Lines~\ref{line:ap} to \ref{line:ap_map} take $O(n)$ time in total during the whole procedure. 
Thus, the time complexity of the whole procedure is $O(m)$. 
This completes the proof. 
\end{proof}

\bibliographystyle{splncs03}
\bibliography{tconn.bib}

\begin{thebibliography}{10}
\providecommand{\url}[1]{\texttt{#1}}
\providecommand{\urlprefix}{URL }

\bibitem{cook1998combinatorial}
Cook, W.J., Cunningham, W.H., Pulleyblank, W.R., Schrijver, A.: Combinatorial
  Optimization. Wiley-Interscience Series in Discrete Mathematics and
  Optimization, John Wiley \& Sons, New York (1998),
  \url{https://onlinelibrary.wiley.com/doi/book/10.1002/9781118033142}

\bibitem{frank1996survey}
Frank, A.: A survey on {$T$}-joins, {$T$}-cuts, and conservative weightings.
  In: Mikl{\'o}s, D., S{\'o}s, V.T., Sz{\H{o}}nyi, T. (eds.) Combinatorics,
  Paul Erd{\H{o}}s is Eighty, Bolyai Society Mathematical Studies, vol.~2, pp.
  213--252. Bolyai Society (1996)

\bibitem{10.1145/3183369}
Gabow, H.N.: Data structures for weighted matching and extensions to
  $b$-matching and $f$-factors. ACM Trans. Algorithms  14(3) (2018)

\bibitem{doi:10.1137/16M1106225}
Gabow, H.N., Sankowski, P.: Algorithms for weighted matching generalizations
  {II}: $f$-factors and the special case of shortest paths. SIAM Journal on
  Computing  50(2),  555--601 (2021)

\bibitem{JUTTNER202434}
J\"uttner, A., Kir\'aly, C., Mendoza-Cadena, L.M., Pap, G., Schlotter, I.,
  Yamaguchi, Y.: Shortest odd paths in undirected graphs with conservative
  weight functions. Discrete Applied Mathematics  357,  34--50 (2024)

\bibitem{korach1992tight}
Korach, E., Penn, M.: Tight integral duality gap in the {C}hinese postman
  problem. Mathematical Programming  55(1),  183--191 (1992)

\bibitem{kv2008}
Korte, B., Vygen, J.: Combinatorial Optimization: Theory and Algorithms,
  Algorithms and Combinatorics, vol.~21. Springer Berlin Heidelberg, 4 edn.
  (2008)

\bibitem{lp1986}
Lov{\'a}sz, L., Plummer, M.D.: Matching Theory, North-Holland Mathematics
  Studies, vol. 121. North-Holland, Amsterdam (1986), also published as Annals
  of Discrete Mathematics, Vol. 29

\bibitem{doi:10.1137/23M158156X}
Schlotter, I., Seb\H{o}, A.: Odd paths, cycles, and \(t\)-joins: Connections
  and algorithms. SIAM Journal on Discrete Mathematics  39(1),  484--504 (2025)

\bibitem{schrijver2003}
Schrijver, A.: Combinatorial Optimization: Polyhedra and Efficiency, Algorithms
  and Combinatorics, vol.~24. Springer Berlin Heidelberg (2003)

\bibitem{sebo1990}
Seb{\H{o}}, A.: Undirected distances and the postman-structure of graphs. J.
  Comb. Theory, Ser. {B}  49(1),  10--39 (1990)

\bibitem{10.1007/3-540-45535-3_30}
Seb{\H{o}}, A., Tannier, E.: Connected joins in graphs. In: Aardal, K.,
  Gerards, B. (eds.) Integer Programming and Combinatorial Optimization. pp.
  383--395. Springer Berlin Heidelberg, Berlin, Heidelberg (2001)

\bibitem{sebHo2004metric}
Seb{\H{o}}, A., Tannier, E.: On metric generators of graphs. Mathematics of
  Operations Research  29(2),  383--393 (2004)

\bibitem{seymour1981odd}
Seymour, P.D.: On odd cuts and plane multicommodity flows. Proceedings of the
  London Mathematical Society  3(1),  178--192 (1981)

\end{thebibliography}

\end{document}